\documentclass[fleqn,usenatbib]{mnras}

\usepackage{newtxtext,newtxmath}
\usepackage{pdflscape}
\usepackage[T1]{fontenc}

\DeclareRobustCommand{\VAN}[3]{#2}
\let\VANthebibliography\thebibliography
\def\thebibliography{\DeclareRobustCommand{\VAN}[3]{##3}\VANthebibliography}

\usepackage{graphicx}	% Including figure files
\usepackage{amsmath}	% Advanced maths commands
\usepackage{longtable}
\usepackage{multicol}
\usepackage{array}
\usepackage{caption}
\title[A theoretical scenario for Gaia RR Lyrae]{A theoretical scenario for Galactic RR Lyrae in the Gaia
  database: constraints on the parallax offset}

\author[M. Marconi et al.]{
M. Marconi,$^{1}$\thanks{E-mail: marcella.marconi@inaf.it}
R. Molinaro,$^{1}$
V. Ripepi,$^{1}$
S. Leccia,$^{1}$
I. Musella,$^{1}$
G. De Somma,$^{1,2,3}$
\newauthor{M. Gatto,$^{1,2}$
and
M. I. Moretti$^{1}$}\\
$^{1}$INAF-Osservatorio Astronomico di Capodimonte, Via Moiariello 16, 80131 Napoli, Italy\\
$^{2}$Dipartimento di Fisica ‘E. Pancini’, Universitá di Napoli ‘Federico II’, Compl. Univ. di Monte S. Angelo, Edificio G, Via Cinthia, I-80126 Napoli, Italy\\
$^{3}$Istituto Nazionale di Fisica Nucleare (INFN)-Sez. di Napoli, Compl. Univ.di Monte S. Angelo, Edificio G, Via Cinthia, I-80126 Napoli, Italy}

\date{Accepted XXX. Received YYY; in original form ZZZ}

\pubyear{2015}

\begin{document}
\label{firstpage}
\pagerange{\pageref{firstpage}--\pageref{lastpage}}
\maketitle

\begin{abstract}
On the basis of an extended set of nonlinear convective RR Lyrae pulsation models we derive the first theoretical light curves in the Gaia bands $G$, $G_{BP}$ and $G_{RP}$ and the corresponding intensity-weighted mean magnitudes and pulsation amplitudes.
The effects of chemical composition on the derived Bailey diagrams in the Gaia filters are discussed for both Fundamental and First Overtone mode pulsators.
The inferred mean magnitudes and colors are used to derive the first theoretical Period-Wesenheit relations for RR Lyrae in the Gaia filters.
The application of the theoretical Period-Wesenheit relations for both the Fundamental and First Overtone mode to  Galactic RR Lyrae in the Gaia Data Release 2 database and complementary information on individual metal abundances, allows us to derive theoretical estimates of their individual parallaxes. These results are compared with the astrometric solutions to conclude that a very small offset, consistent with zero, is required in order to reconcile the predicted distances with Gaia results. 
\end{abstract}


%%%%%%%%%%%%%%%%%%%%%%%%%%%%%%%%%%%%%%%%%%%%%%%%%%

%%%%%%%%%%%%%%%%% BODY OF PAPER %%%%%%%%%%%%%%%%%%

\section{Introduction}

RR Lyrae are old low mass stars that, during the central
Helium burning phase, show mainly radial pulsation while crossing
the classical instability strip in the Color-Magnitude diagram.
From the observational point of view, they represent the most numerous
class of pulsating stars in the Milky Way and, being associated to old
stellar populations, are typically found in globular cluster and
abundant in the Galactic halo and bulge.
The investigation of RR Lyrae properties is motivated by their
important role both as distance indicators and tracers of old stellar
populations.
In particular, evolving through the central Helium burning phase, they represent the low mass, Population II counterparts of Classical
Cepheids, as powerful standard candles and calibrators of secondary
distance indicators. In particular, they can be safely adopted  to infer distances to Galactic
globular clusters \citep[see e.g.][and references
therein]{Coppola11,Braga16,Braga18}, the Galactic center \citep[see
e.g.][]{contr18,marconi18b,Griv19}  and Milky Way satellite galaxies \citep[see e.g.][and
references therein]{Coppola15,MV19,VIVAS19}. Being associated to old stellar populations, they represent the basis of an alternative Population II distance scale \citep[see e.g.][to the traditionally adopted Classical Cepheids]{Beaton2016}, more suitable to calibrate secondary distance indicators that are not specifically associated to spiral galaxies \citep[e.g. the Globular Cluster Luminosity Function, see][and references therein]{DiCriscienzo2006}.
The properties that make RR Lyrae standard candles are: i) the well known
relation connecting the absolute visual magnitude $M_V$ to the metal
abundance ${\rm [Fe/H]}$ \citep[see e.g.][and references
therein]{sand93,caputo00,caccclem03,catelan04,dicrisci04,federici12,marconi12,
  marconi15, marconi18a, muraveva18}; ii) the
Period-Luminosity relation in the Near-Infrared (NIR) filters and in
particular in the K 2.2 $\mu{m}$ band \citep[see e.g.][and references
therein]{longmore90,bono03,dallora06,
  Coppola11,ripepi12,Coppola15,marconi15,muraveva15,Braga18,marconi18a}.
In spite of the well-known advantage of using NIR filters \citep[see
e.g.][and references therein]{marconi12,Coppola15}, in the last decades there has been a debate on the coefficient of the metallicity term of the K Band PL relation \citep[see e.g.][and references therein]{bono03,Sollima06,marconi15}. On the other hand, it is interesting to note that many recent determinations \citep[see e.g.][]{sesar17,muraveva18} seem to converge towards the predicted coefficient by Marconi et al. (2015), with values in the range 0.16-0.18 mag/dex. As for the optical bands, our recently developed
theoretical scenario \citep{marconi15} showed that, apart from the
$M_V -  {\rm [Fe/H]}$ relation that is affected by a number of uncertainties \citep[e.g. a possible nonlinearity, the metallicity scale with the
associated $\alpha$ elements enhancement and helium abundance
variations, as well as 
evolutionary effects, see][for a discussion]{caputo00,marconi18a},  the metal-dependent Period-Wesenheit (PW) relations are predicted to be sound tools to infer individual distances. In particular, for
the B,V band combination, \citet{marconi15} demonstrated that the
inferred PW relation is independent of metallicity. 
In order to test this theoretical tool, we need to compare the predicted individual distances with independent reliable distance estimates, as for example the astrometric ones recently obtained by the Gaia satellite \citep{Gaia2016}.
To this purpose, in the present paper we transform the predicted light curves derived for RR Lyrae models with a wide range of
chemical compositions \citep{marconi15,marconi18a} into the Gaia bands, derive the first theoretical PW relations in these filters and
apply them to Gaia Data Release 2 Database \citep[hereinafter Gaia DR2][]{Gaia2018,Clementini2019,Ripepi2019}.
The organization of the paper is detailed in the following. In Section 2 we
summarize the adopted theoretical scenario, while in Section 3 we
present the first theoretical light curves in the Gaia filters. From
the inferred mean magnitudes and colors the new theoretical PW
relations are derived in Section 4, that also includes a discussion of the effects of
variations in the input chemical abundances.
In Section 5 the obtained relations are applied to Gaia Galactic
RR Lyrae with available periods, parallaxes and mean magnitudes to infer independent predictions
on their individual parallaxes, to be compared with Gaia DR2 results.
%Finally, in Section 6, Gaia DR2 parallaxes are used in combination with the metal
%dependent theoretical PW relations to infer a metallicity distribution
%of the investigated sample of Galactic RR Lyrae.
The conclusions close the paper.

\section{The theoretical scenario}

In two recent papers \citep{marconi15,marconi18a}, we presented an
updated theoretical scenario for RR Lyrae stars.
In \citet{marconi15} a wide range of chemical abundances was
considered, spanning from typical metal-poor globular clusters values
($Z=0.0001$)
to typical abundances of Galactic Disk and Bulge RR Lyrae  ($Z=0.02$), with a standard helium content ranging from $Y=0.245$ for the most metal poor abundances to $Y=0.28$ at solar metallicity \citep[see][for details]{marconi15}.
Moreover, two stellar masses and three luminosity levels were adopted  for each selected
chemical composition, in order to take into account not only RR Lyrae
located on the Zero Age Horizontal Branch (ZAHB) but also evolved objects \citep[see][for details]{marconi15}.
In order to take into account possible variations in the helium
abundance,  in \citet{marconi18a} we recomputed the model sets
presented in \citet{marconi15}  by increasing the helium content  to $Y=0.30$ and $Y=0.40$ \citep[see][for details]{marconi18a,marconi18b}.
For each selected combination of $Z$, $Y$, stellar
mass and luminosity, the system of nonlinear hydrodinamical and
convective equations was integrated till a stable limit cycle is
achieved in the Fundamental (F) or First Overtone (FO) mode.
The resulting bolometric light and radial velocity curves represent
an extended dataset of theoretical templates allowing the
investigation of the effect of both $Z$ and $Y$ not only on the
pulsation period but also on the amplitude and the morphology of the
luminosity and radial velocity variations. A similar analysis can be
performed with radius curves and the variations of all other relevant
quantities, e.g. gravity and temperature, along a pulsation cycle.
A subset of the produced theoretical atlas, for $Z=0.001$, $Y=0.245$,
$M=0.64M_{\odot}$ and $\log{L=1.67}$,  is shown in Figures 1 and 2
for F  and FO models, respectively.
In each left panel the bolometric light curve is plotted for the
labelled effective temperature, whereas the corresponding radial
velocity is shown in the right panel, with the labelled
period value.

\begin{figure}
\begin{multicols}{2}
\vbox{
    \includegraphics[width=8cm]{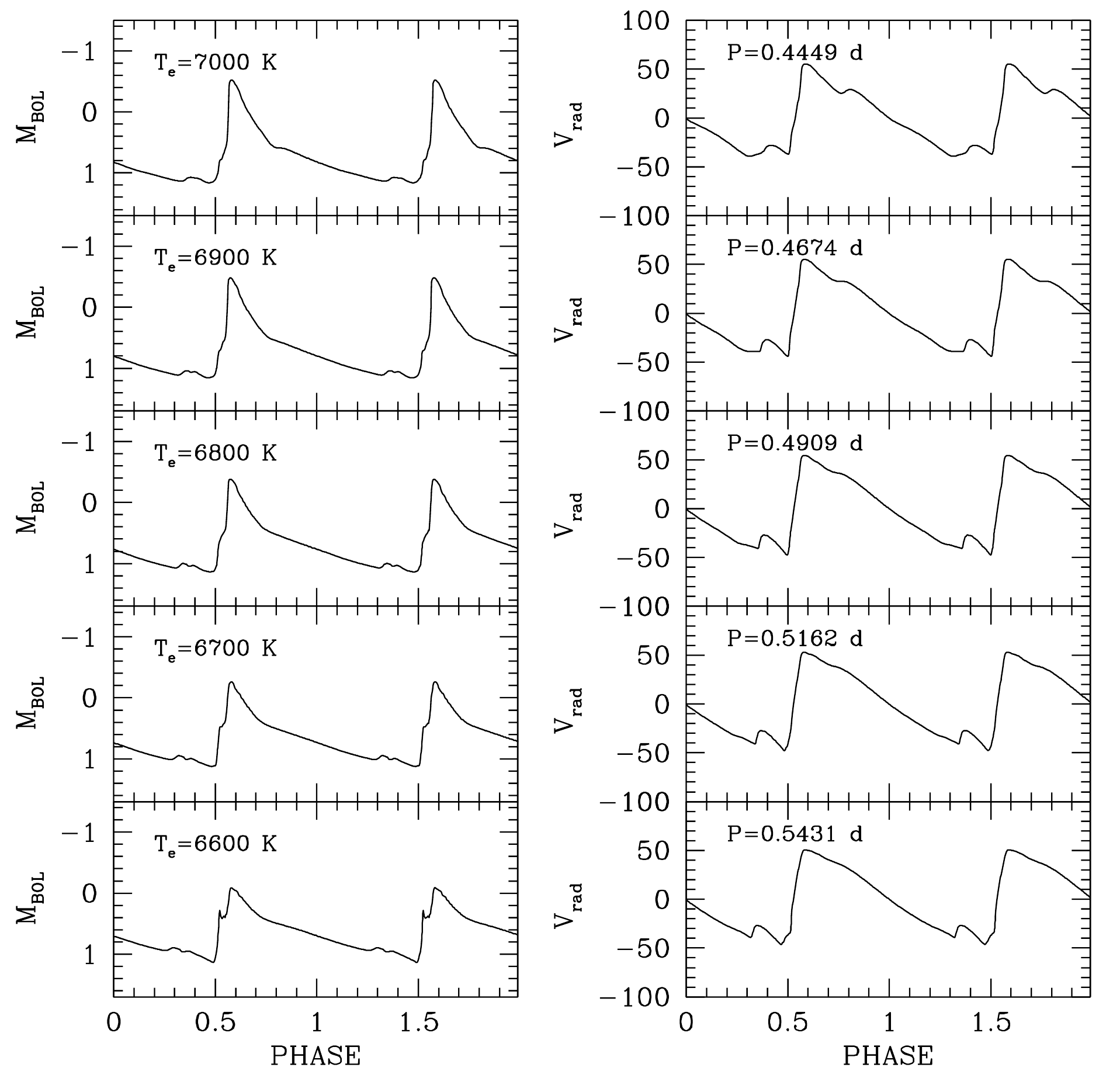}\par 
    \includegraphics[width=8cm]{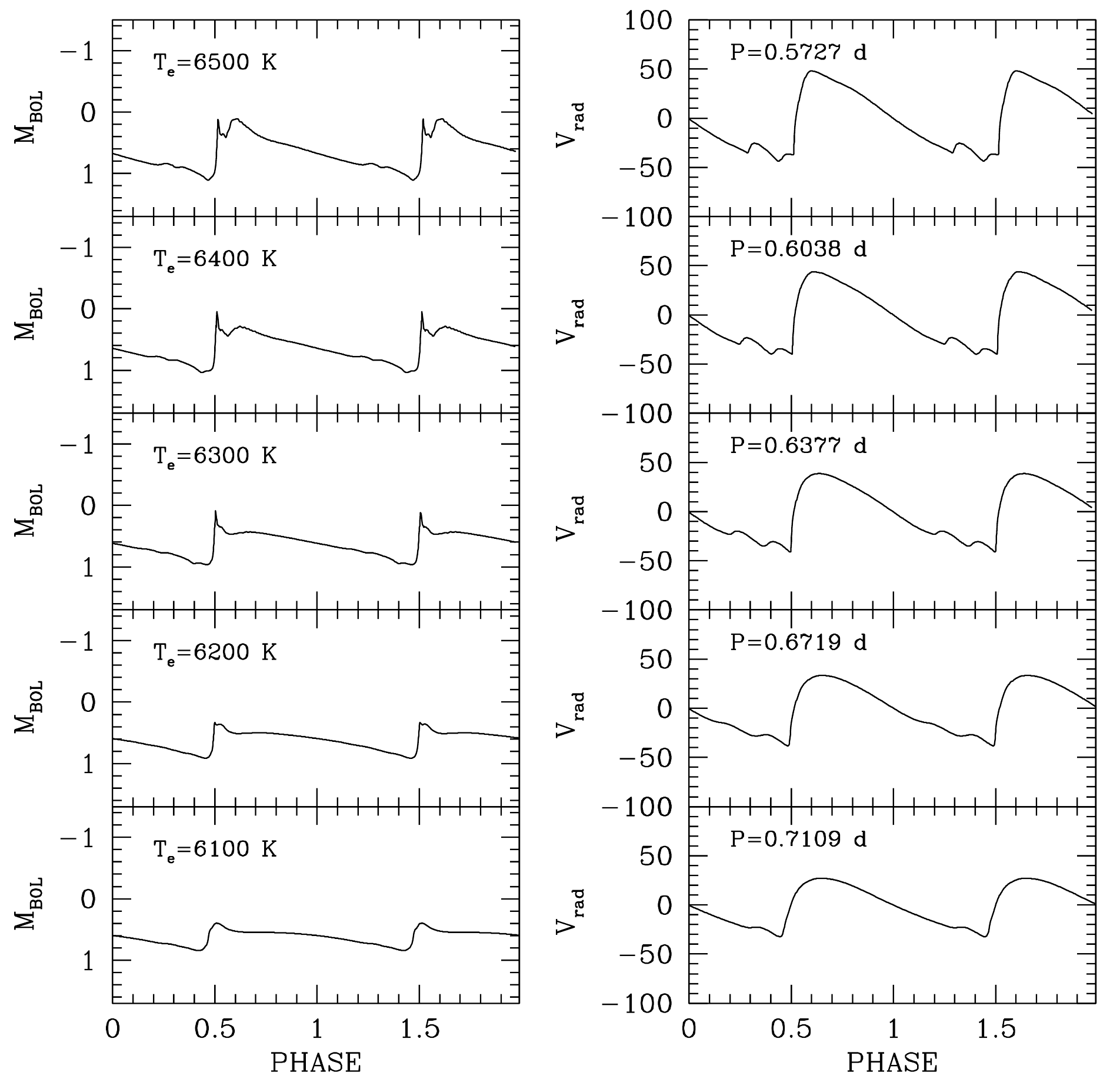}\par 
    }
\end{multicols}
\caption{\label{fig1} Predicted bolometric light curves for F-mode
  RR Lyrae models assuming  $Z=0.001$, $Y=0.245$,
$M=0.64M_{\odot}$ and $\log{L=1.67}$. In each left panel the bolometric light curve is plotted for the
labelled effective temperature, whereas the radial
velocity curve is shown in the corresponding right panel, with the labelled
period value.}
\end{figure}

\begin{figure}
\includegraphics[width=\columnwidth]{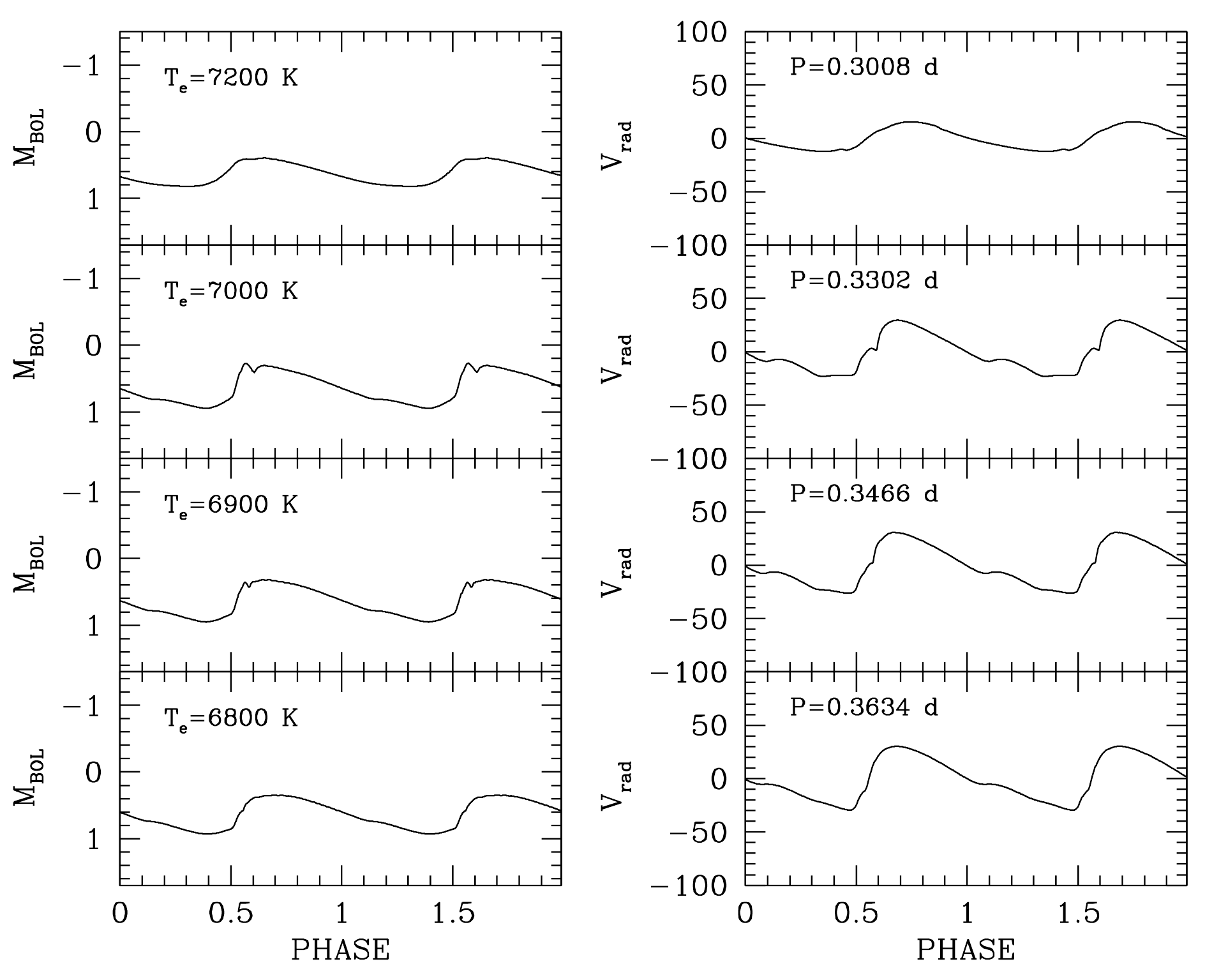}
\caption{The same as Figure 1 but for FO-mode RR Lyrae models at the
  same chemical composition, mass and luminosity.}
\label{fig2}
\end{figure}

These plots show an example of theoretical bolometric magnitude and
radial velocity variations that are made available to the community
upon request as based on the models computed in \citet{marconi15,marconi18a}

\section{Theoretical light curves for RR Lyrae in the Gaia filters}

The predicted bolometric light curves discussed above have been transformed into the
Gaia bands, namely $G$, $G_{BP}$ and $G_{RP}$, by using the Bolometric Corrections (BC) tables provided by \citet{che19}. This database is based on the most recent and adopted spectral libraries and covers a wide variety of photometric systems, including the Gaia passbands \citep[see][for all the details]{che19}. Moreover, these authors provide BC tables for different elemental composition values covering the range considered in this work. By adopting the effective temperature $T_e$ and the gravity $\log(g)$ model curves as input, we  used a proprietary C code to interpolate the BC tables. When the chemical composition of our models coincides with one specific value of \citet{che19} grid, we select one BC table and a bi-linear interpolation is performed along the $T_e-\log(g)$ direction. On the other hand, if the chemical composition of our models falls between two of the quoted BC tables, we first apply our routine on each neighbouring table, and then interpolate linearly between the two metallicity values. For this procedure we used $M_{bol}^\odot=4.79$ mag, consistently with the value adopted in the pulsation code.
If the most recent IAU accepted value of the sun bolometric magnitude (4.74 mag) were assumed instead of the adopted 4.79 mag, we would obtain differences in the predicted individual mean magnitudes of the order of 0.01-0.02 mag.
In Figures from 3 to 9 we show the first theoretical RR Lyrae
light curves transformed into the Gaia bands changing the metallicity (from
$Z=0.0001$ to $Z=0.02$, see captions) and considering both F (top panels)
and FO (bottom panels) models. The stellar parameter selection is the same  as in \citet{marconi15}.

\begin{figure}
%\begin{multicols}{3}
\vbox{
    \includegraphics[width=8cm]{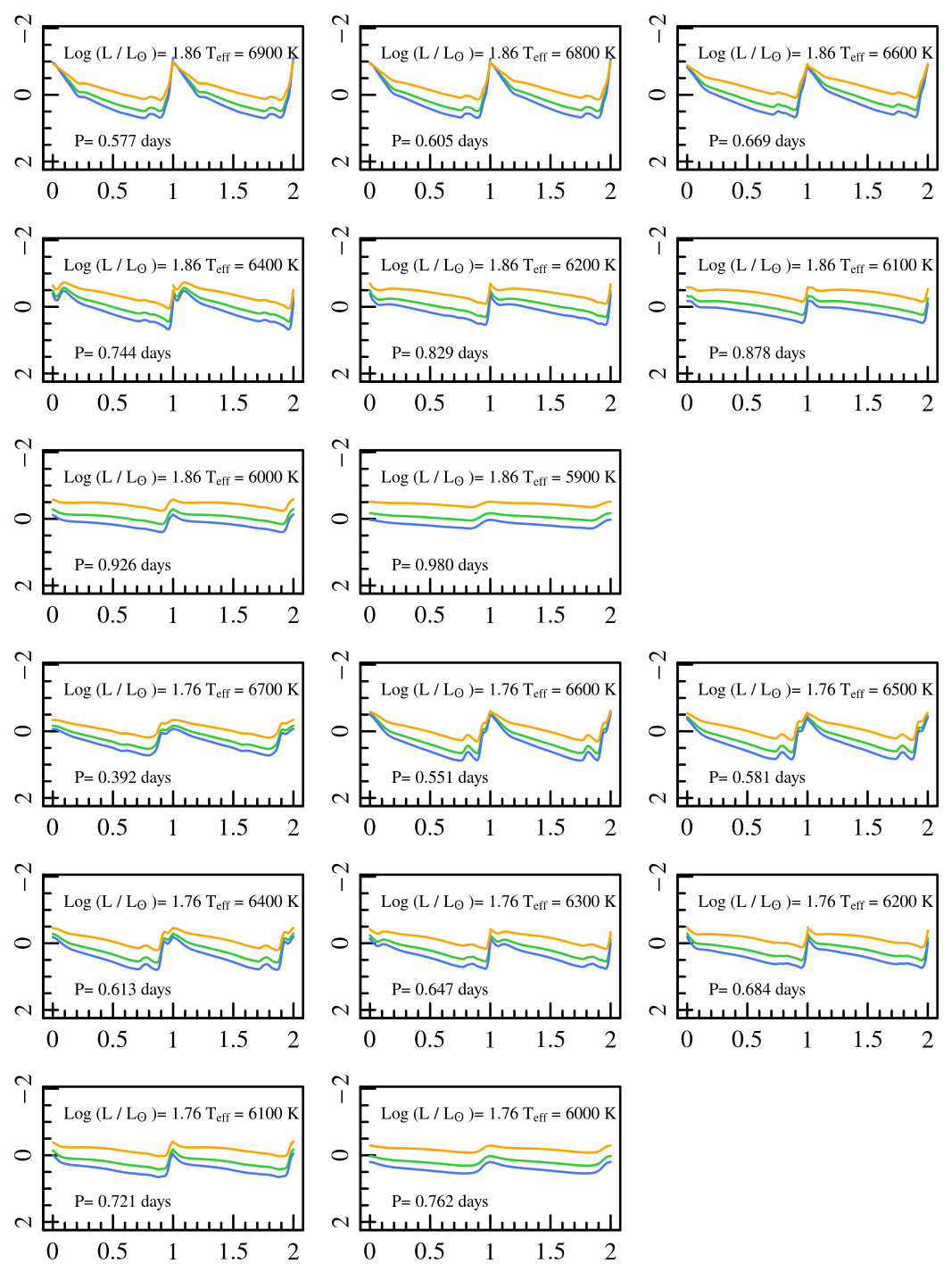}\par 
        \vspace{1cm}
    \includegraphics[width=8cm]{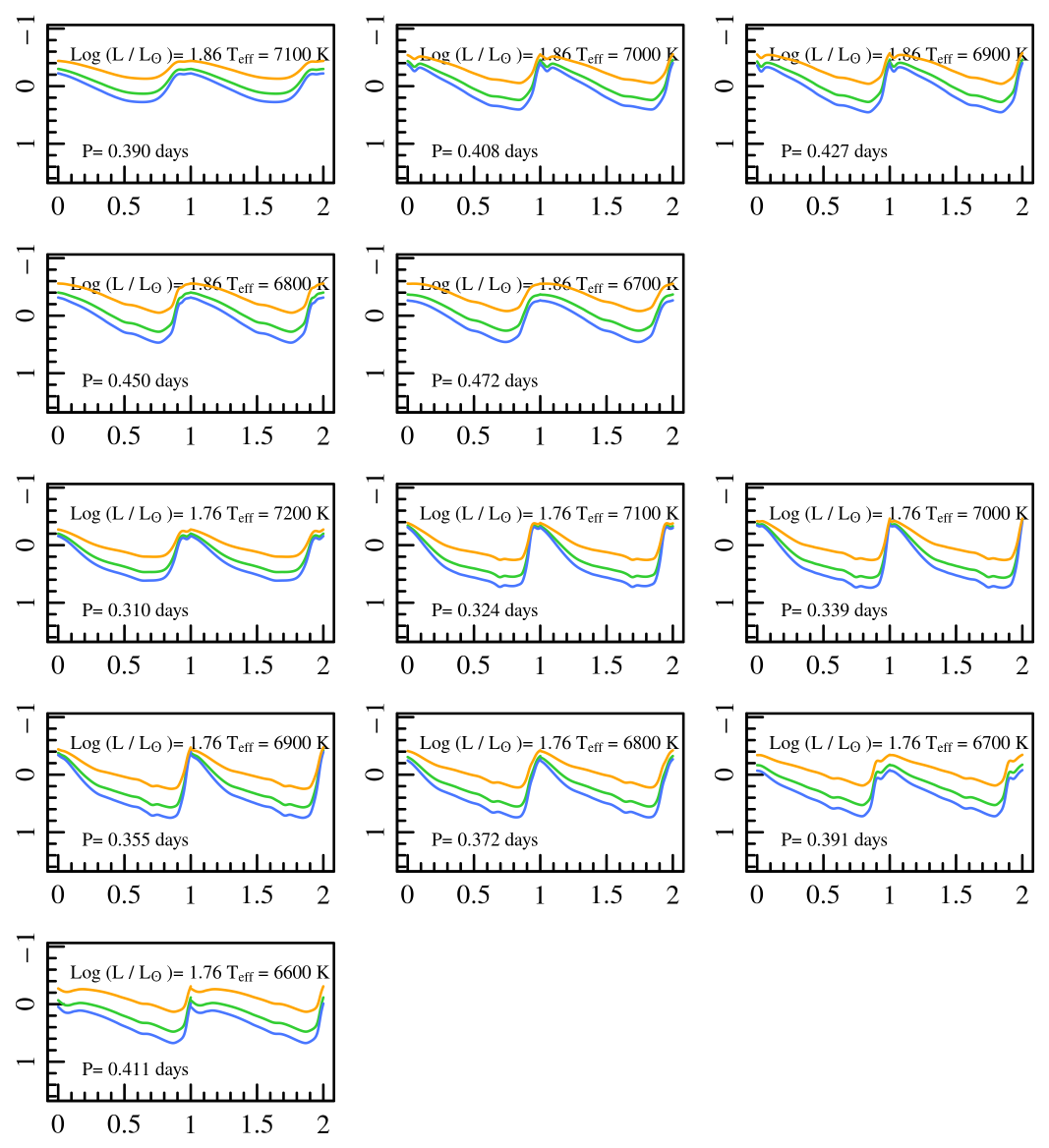}\par 
    }
%\end{multicols}
\caption{\label{fig1} Theoretical light curves transformed into the Gaia filters $G_{BP}$ (orange), $G$ (green) and $G_{RP}$ (blue) for F-mode (top) and FO-mode (bottom)
  RR Lyrae models assuming  $Z=0.0001$, $Y=0.245$ and $M=0.80M_{\odot}$.}
\end{figure}

\begin{figure}
%\begin{multicols}{3}
\vbox{
    \includegraphics[width=8cm]{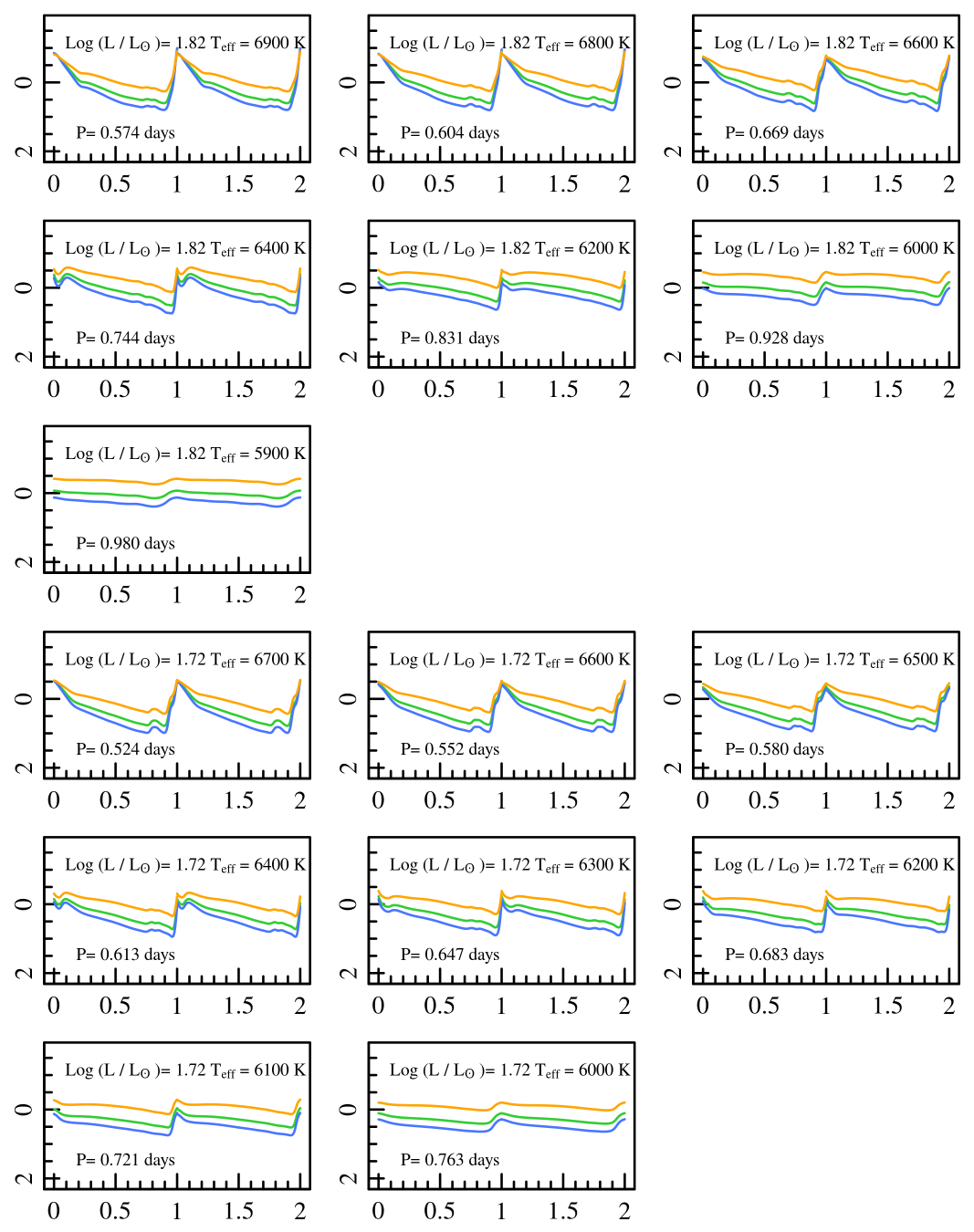}\par 
        \vspace{1cm}
    \includegraphics[width=8cm]{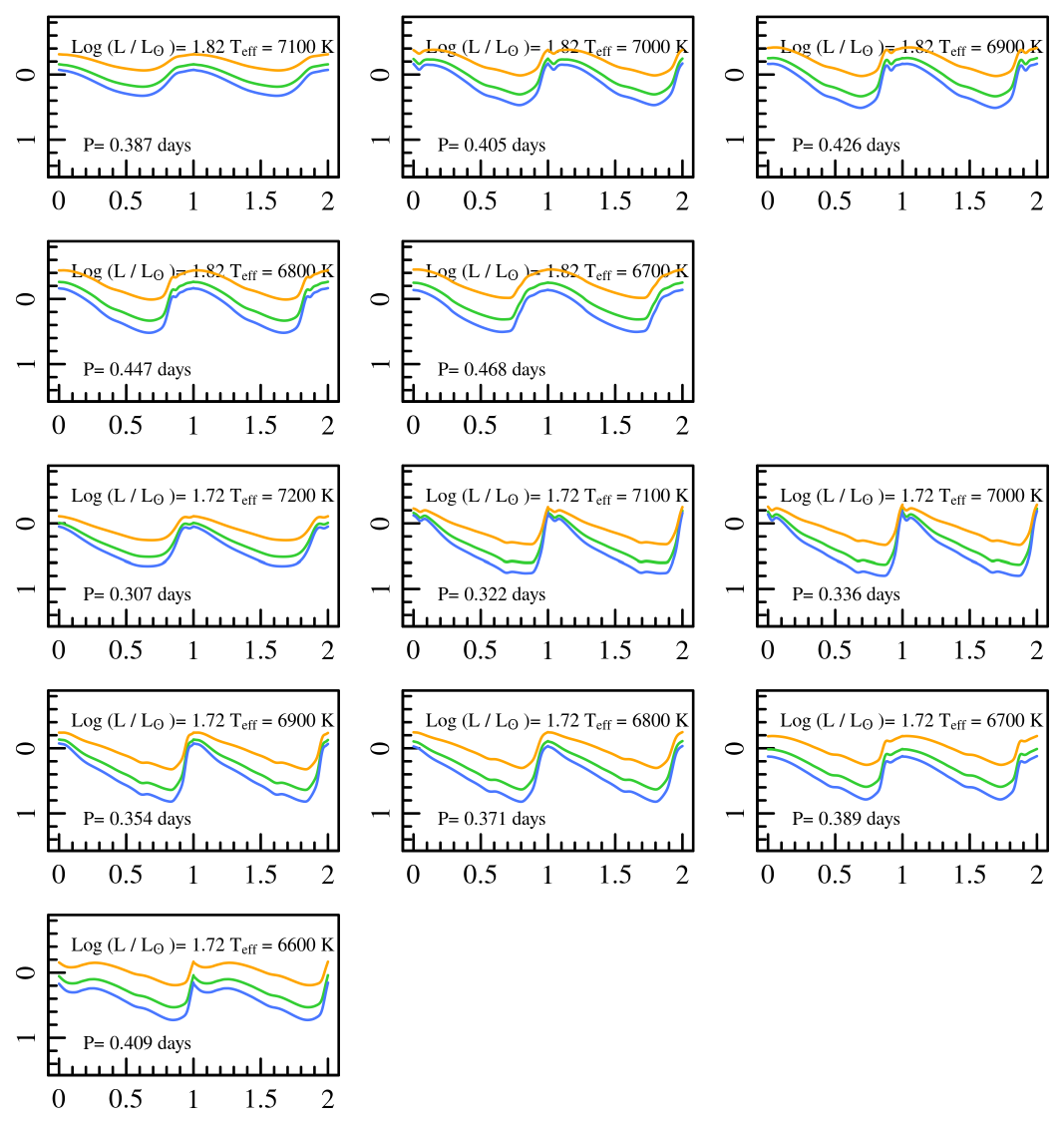}\par 
    }
%\end{multicols}
\caption{\label{fig1} The same as in Figure 3 but assuming  $Z=0.0003$, $Y=0.245$ and $M=0.716M_{\odot}$.}
\end{figure}

\begin{figure}
%\begin{multicols}{3}
\vbox{
    \includegraphics[width=8cm]{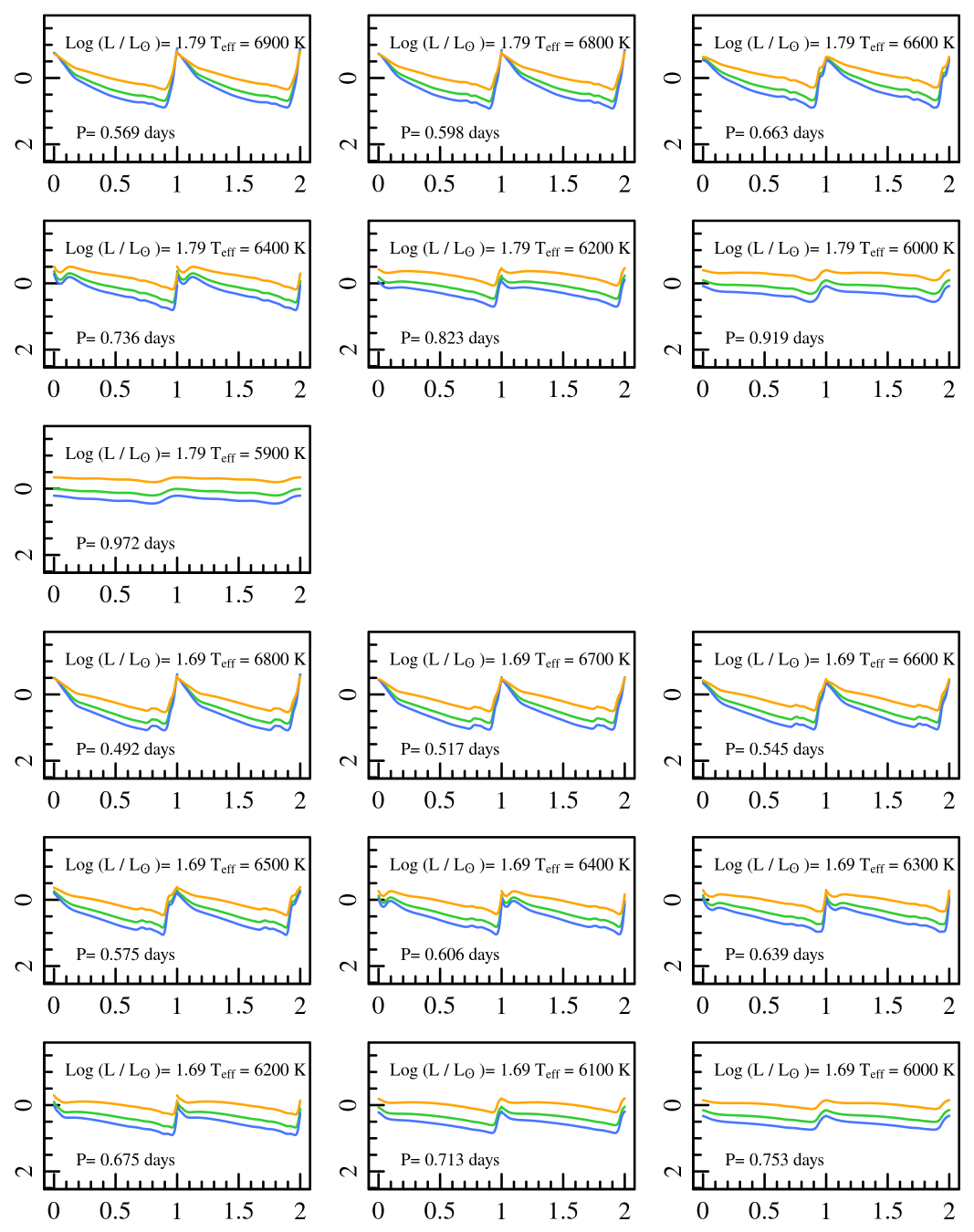}\par 
        \vspace{1cm}
    \includegraphics[width=8cm]{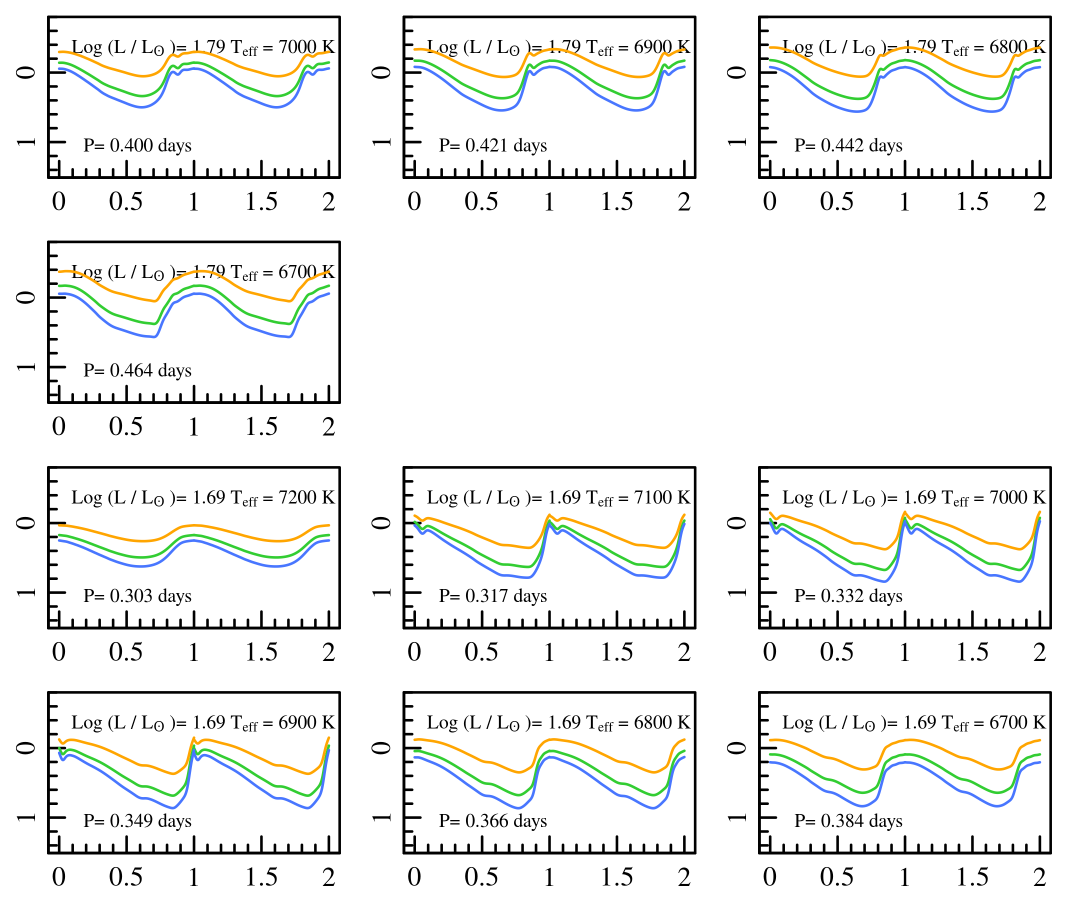}\par 
    }
%\end{multicols}
\caption{\label{fig1} The same as in Figure 3 but assuming  $Z=0.0006$, $Y=0.245$ and $M=0.67M_{\odot}$.}
\end{figure}

\begin{figure}
%\begin{multicols}{3}
\vbox{
    \includegraphics[width=8cm]{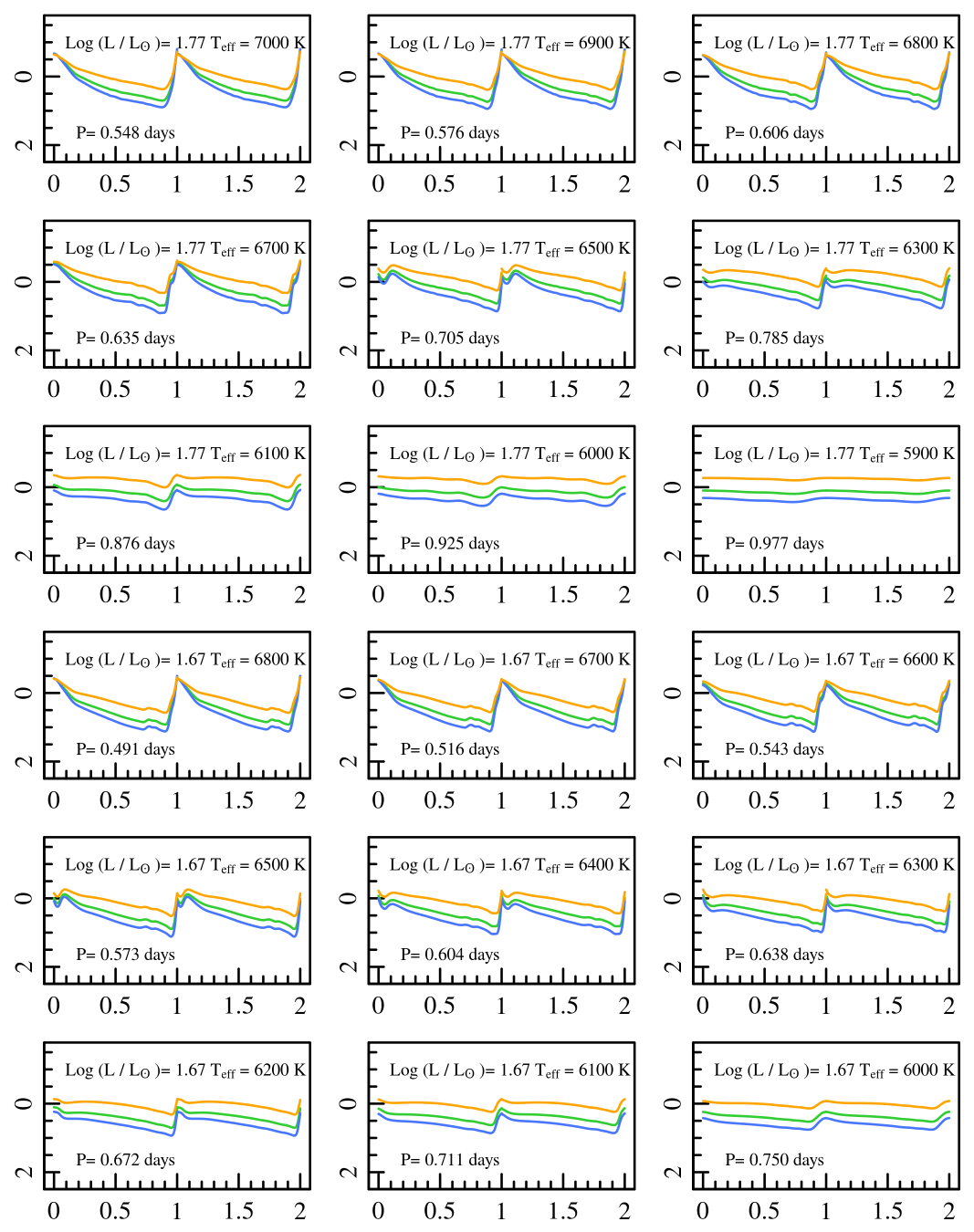}\par 
        \vspace{1cm}
    \includegraphics[width=8cm]{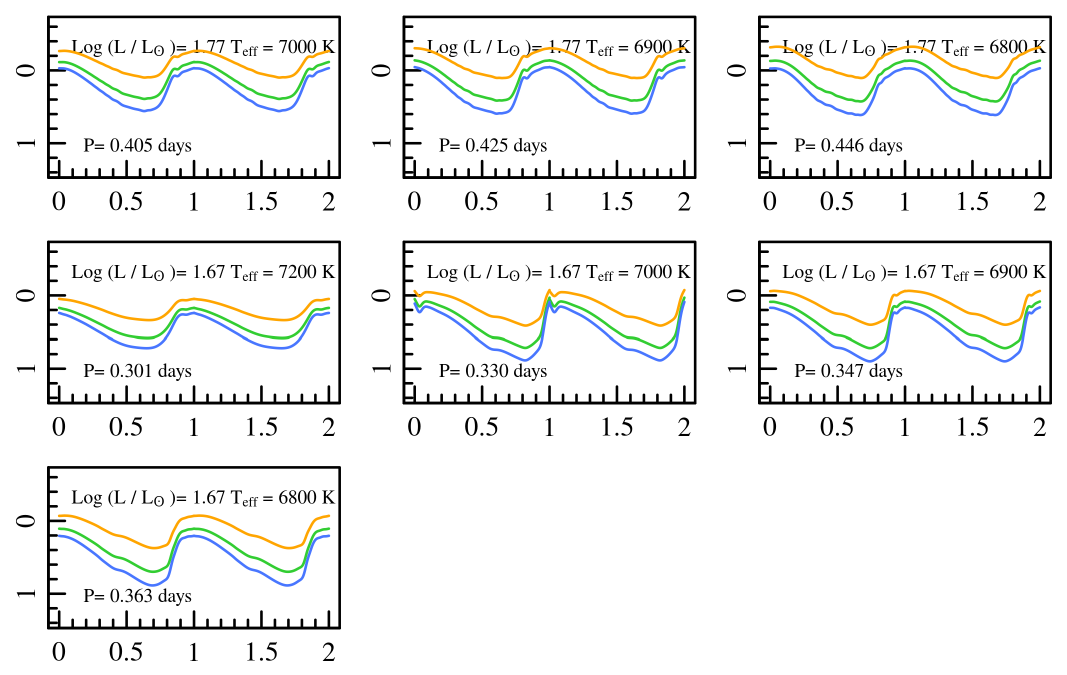}\par 
    }
%\end{multicols}
\caption{\label{fig1} The same as in Figure 3 but assuming  $Z=0.001$, $Y=0.245$ and $M=0.64M_{\odot}$.}
\end{figure}

\begin{figure}
%\begin{multicols}{3}
\vbox{
    \includegraphics[width=8cm]{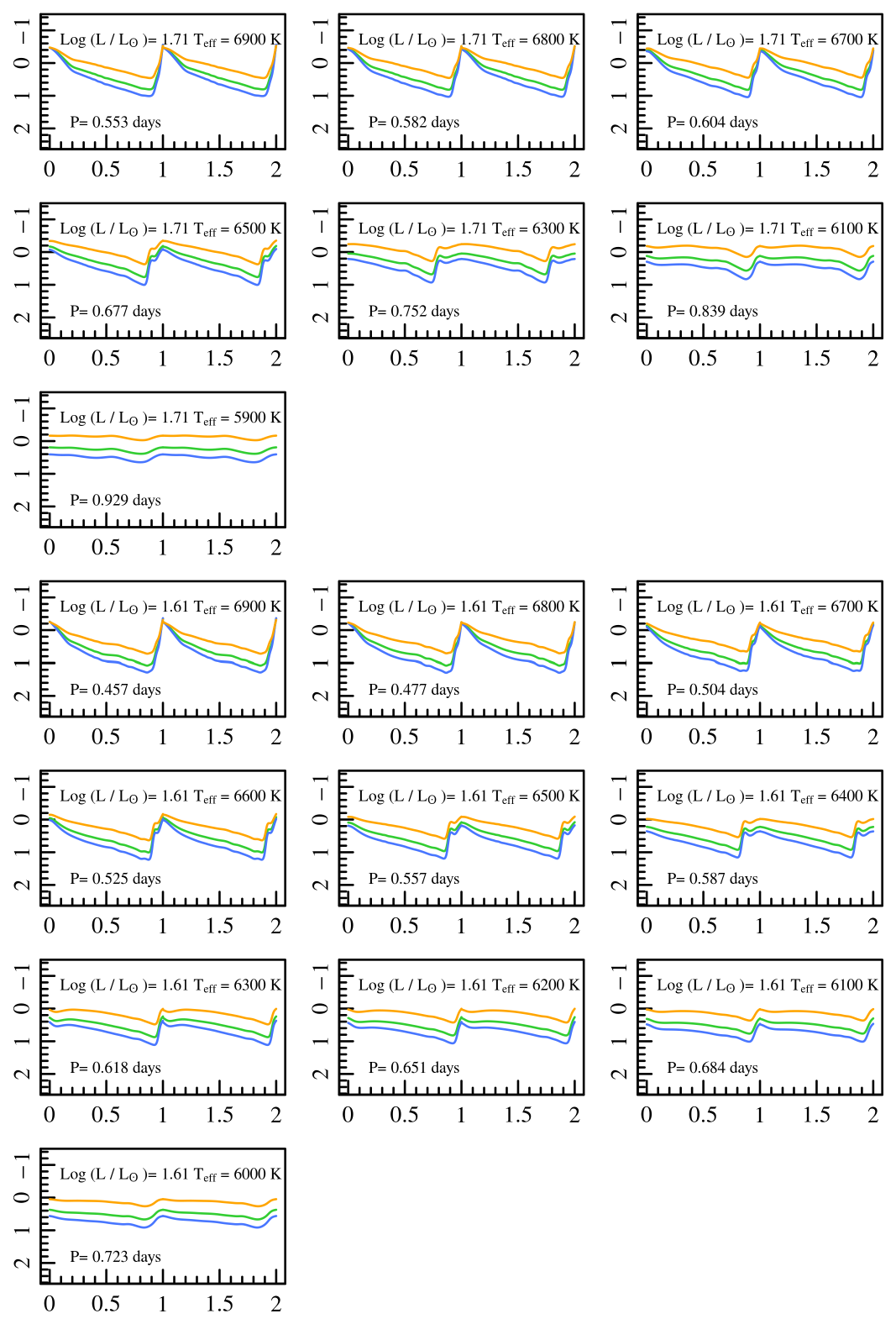}\par 
        \vspace{1cm}
    \includegraphics[width=8cm]{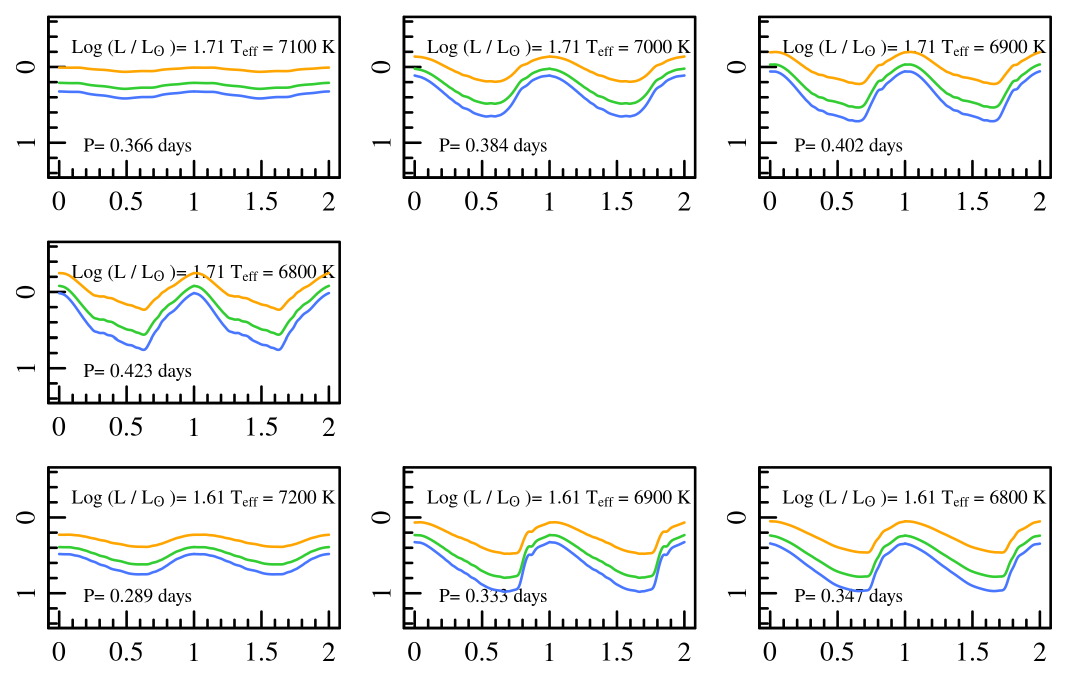}\par 
    }
%\end{multicols}
\caption{\label{fig1} The same as in Figure 3 but assuming  $Z=0.004$, $Y=0.25$ and $M=0.59M_{\odot}$.}
\end{figure}

\begin{figure}
%\begin{multicols}{3}
\vbox{
    \includegraphics[width=8cm]{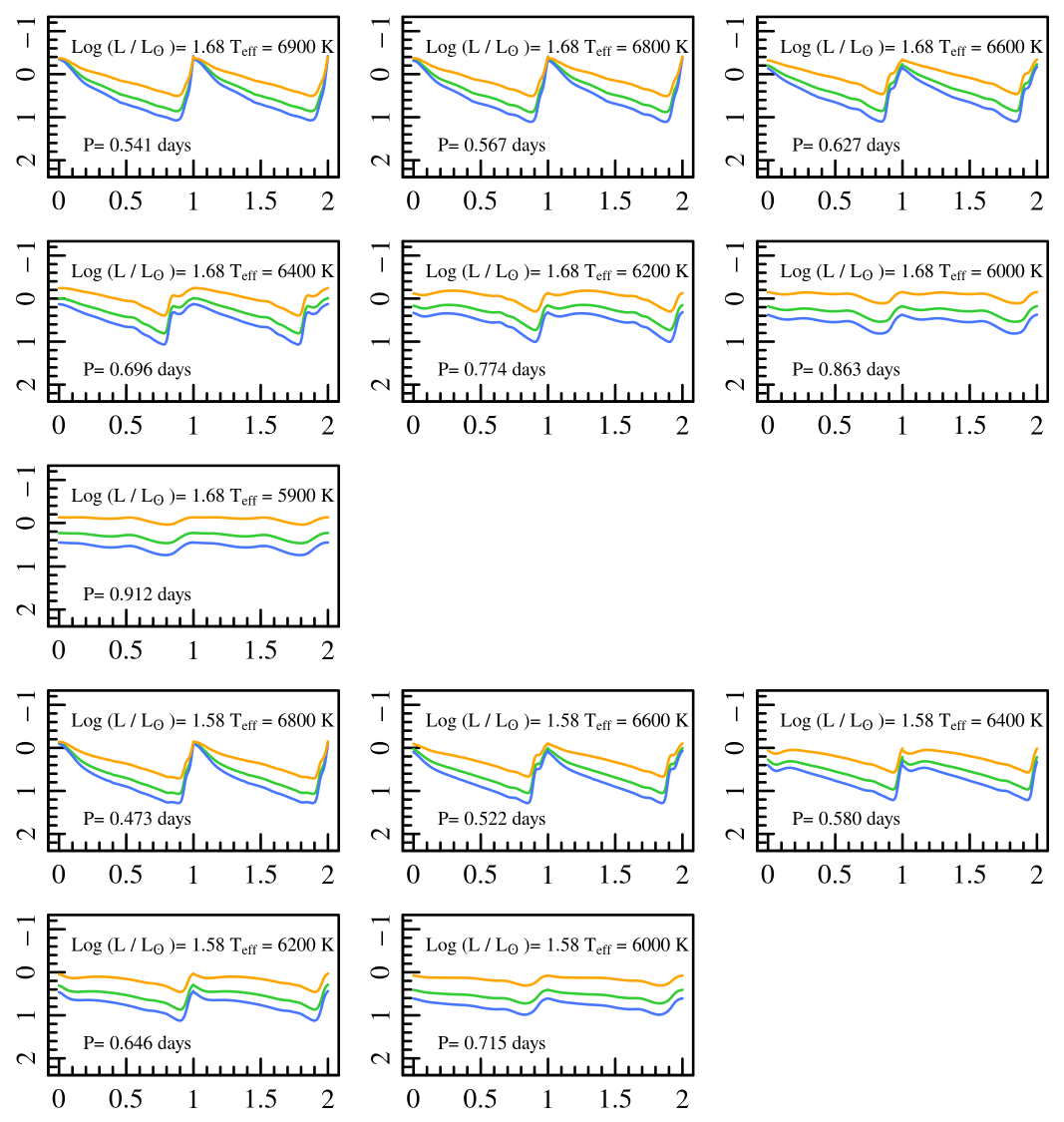}\par 
        \vspace{1cm}
    \includegraphics[width=8cm]{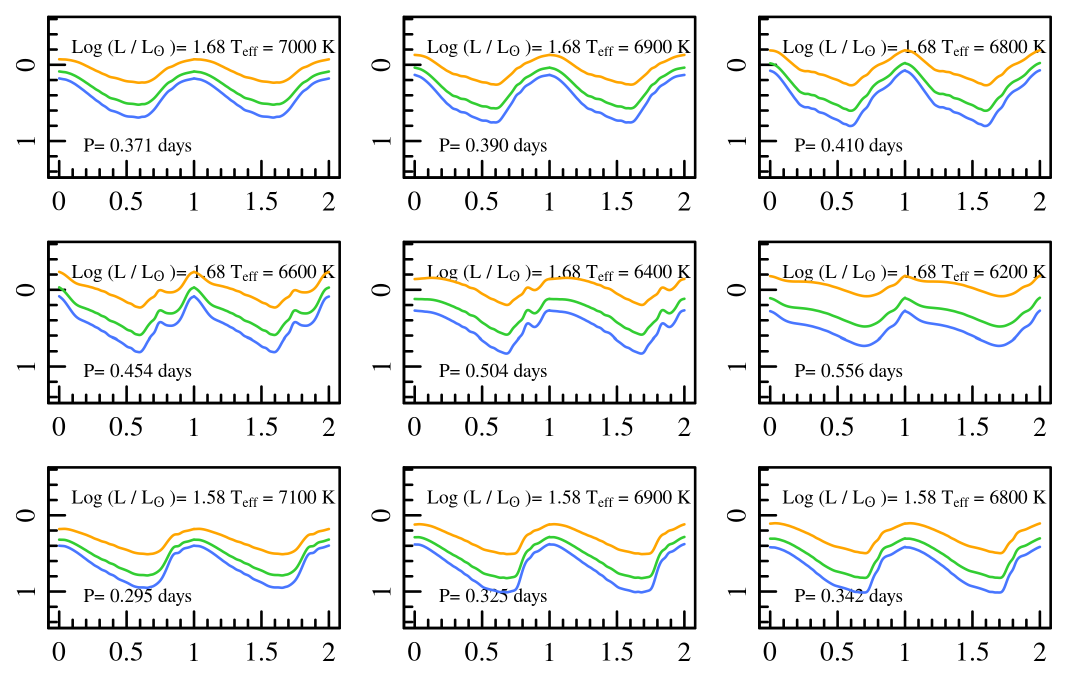}\par 
    }
%\end{multicols}
\caption{\label{fig1} The same as in Figure 3 but assuming  $Z=0.008$, $Y=0.256$ and $M=0.57M_{\odot}$.}
\end{figure}

\begin{figure}
%\begin{multicols}{3}
\vbox{
    \includegraphics[width=8cm]{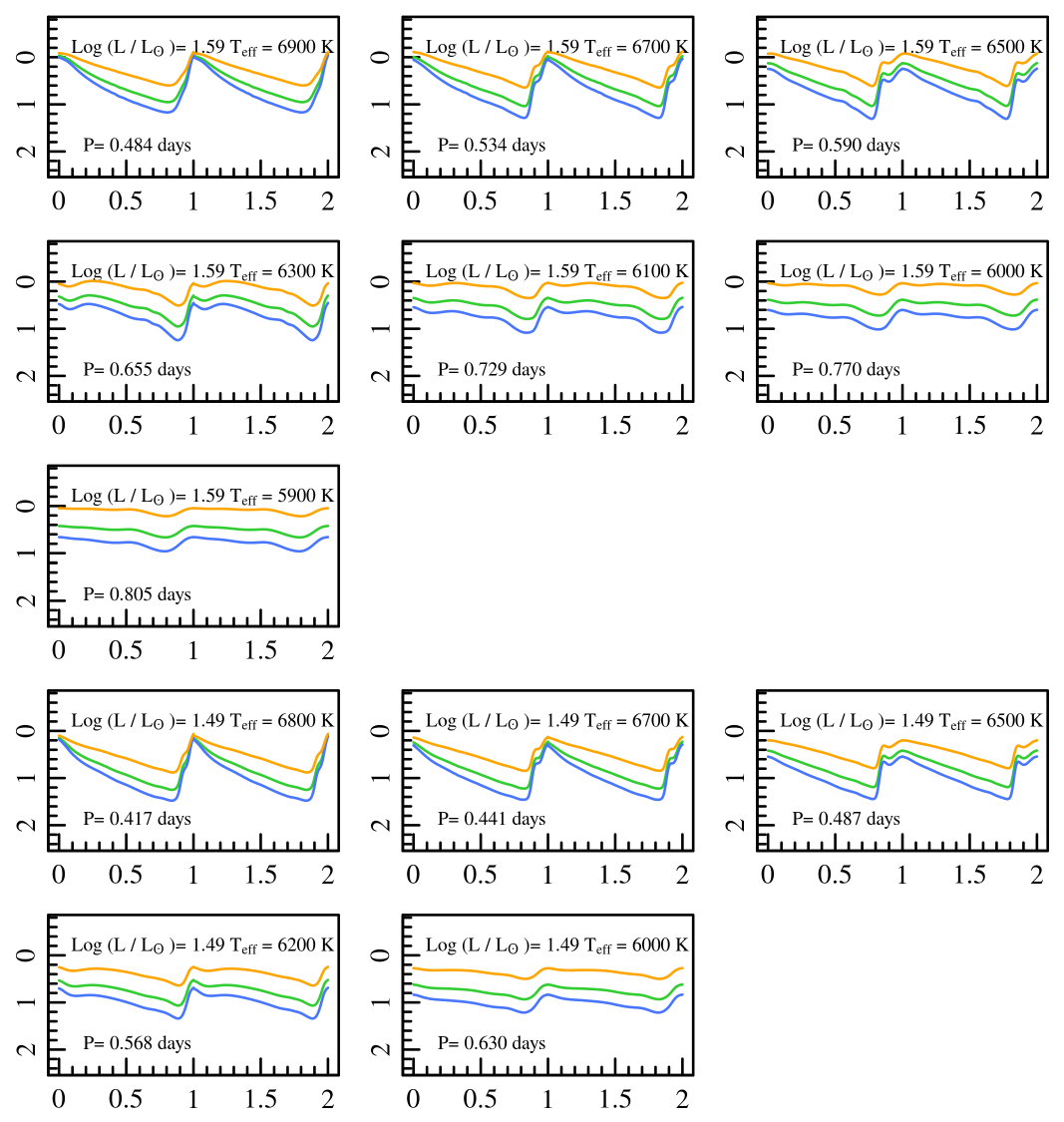}\par
    \vspace{1cm}
    \includegraphics[width=8cm]{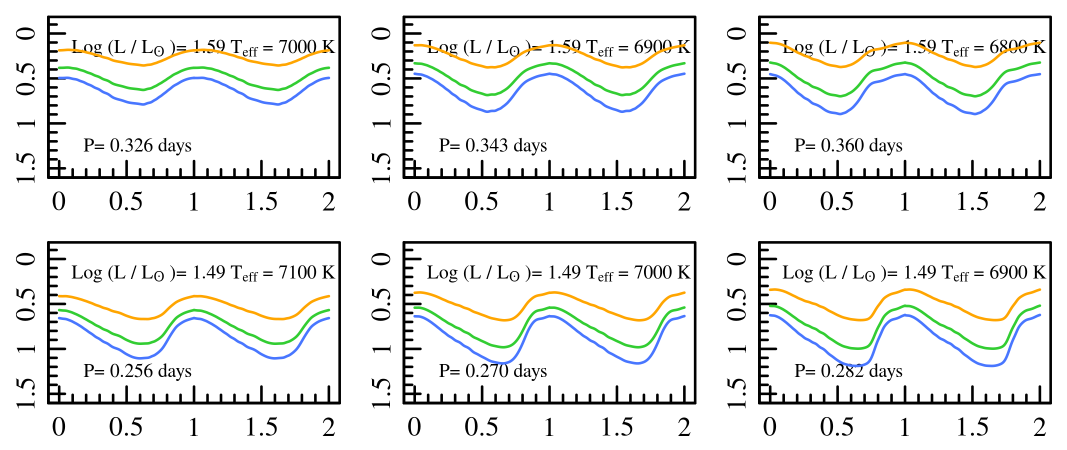}\par 
    }
%\end{multicols}
\caption{\label{fig1} The same as in Figure 3 but assuming  $Z=0.02$, $Y=0.27$ and $M=0.54M_{\odot}$.}
\end{figure}

Similar plots but varying the
helium abundance, up to 0.30 and 0.40, for the stellar masses and
luminosities as in \citet{marconi18a} are available upon request to the authors.
These plots show that the amplitude and morphology of the light curves
in the Gaia bands 
follow the same trends with the effective temperature  as
in the optical bands \citep[see e.g.][and references therein]{marconi15}.
In particular, we notice that:
\begin{enumerate}
 \item At fixed mass and luminosity, the pulsation amplitude of F
   light curves generally decrease as the
   effective temperature decreases and the pulsation period
   increases (see also Section 4 for more details).
   
  \item First overtone amplitudes do not show a linear behaviour with
    the pulsation period, as they reach a maximum towards the center
    of the FO instability strip to decraese again towards the red
    edge \citep[bell shape, see][for details]{BCCM1997}.

    \item The morphology of F light curves is much more complicated
      than for FO models, with the presence of bumps and dips related
      to the coupling between pulsation and convection in the
      pulsating envelope \citep[see e.g. discussion
      in][]{BS1994,BCCM1997,dicrisci04}.

 \end{enumerate}

 The data points for the plotted theoretical $G$, $G_{BP}$ and $G_{RP}$  light curves are again available upon
 request. These  are used to infer mean magnitudes and colors as well
 as pulsation amplitudes, as discussed in the following.

\subsection{The Bailey Diagram}

On the basis of the transformed light curves discussed above, we are able to build the first predicted Bailey diagram in the
three Gaia bands, varying both $Z$ and $Y$.
In Figure 10 we show the $G_{BP}$ (top panel), $G$
(middle panel) and $G_{RP}$ (lower panel) pulsation amplitudes as a
function of the pulsation period for the labelled metallicities, the corresponding predicted ZAHB masses \citep[see][for details]{marconi15}, namely 0.80$M_{\odot}$ for Z=0.0001,  0.64$M_{\odot}$ for Z=0.001 and 0.57$M_{\odot}$ for Z=0.008, standard $Y$ as in \citet{marconi15} and two luminosity levels corresponding to the ZAHB level (solid lines) and a brighter luminosity by 0.1 dex  (evolved stage, dotted lines), for each fixed mass.
We notice that, in agreement with previous empirical and theoretical results in the optical and near-infrared filters, the following trends can be seen:
\begin{itemize}
    \item the pulsation amplitudes decrease as the band central  wavelength  and the metallicity increase;
    \item an increase in the luminosity level produces a period shift towards longer values. On this basis, in the case of FO RR Lyrae, the location of the described bell-shape in the Bailey diagram can be used to constrain the luminosity level \citep[see e.g.][]{BCCM1997}. 
\end{itemize}

\begin{figure}
%\begin{multicols}{3}
    \includegraphics[width=8cm]{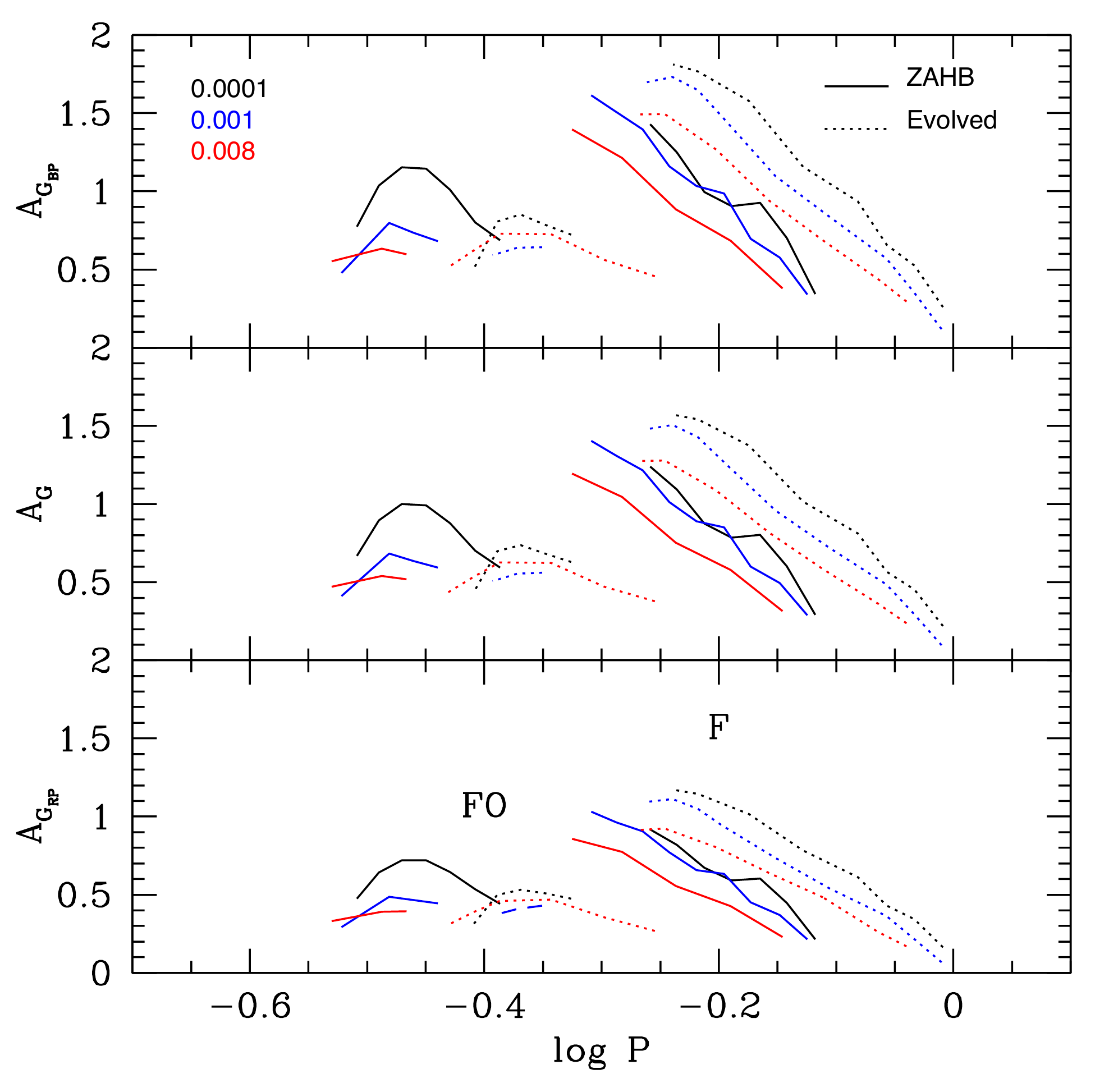}\par 
%\end{multicols}
\caption{\label{fig1} The Bailey diagram in the $G_{BP}$ (top panel), $G$
(middle panel) and $G_{RP}$ (lower panel) filters for the labelled metal abundances, the corresponding predicted ZAHB masses \citep[see][for details]{marconi15}, namely 0.80$M_{\odot}$ for Z=0.0001,  0.64$M_{\odot}$ for Z=0.001 and 0.57$M_{\odot}$ for Z=0.008, standard $Y$ as in \citet{marconi15} and two luminosity levels corresponding to the ZAHB level (solid line) and a brighter luminosity by 0.1 dex (dotted line), for each fixed mass.}
\end{figure}

\begin{figure}
%\begin{multicols}{3}
    \includegraphics[width=8cm]{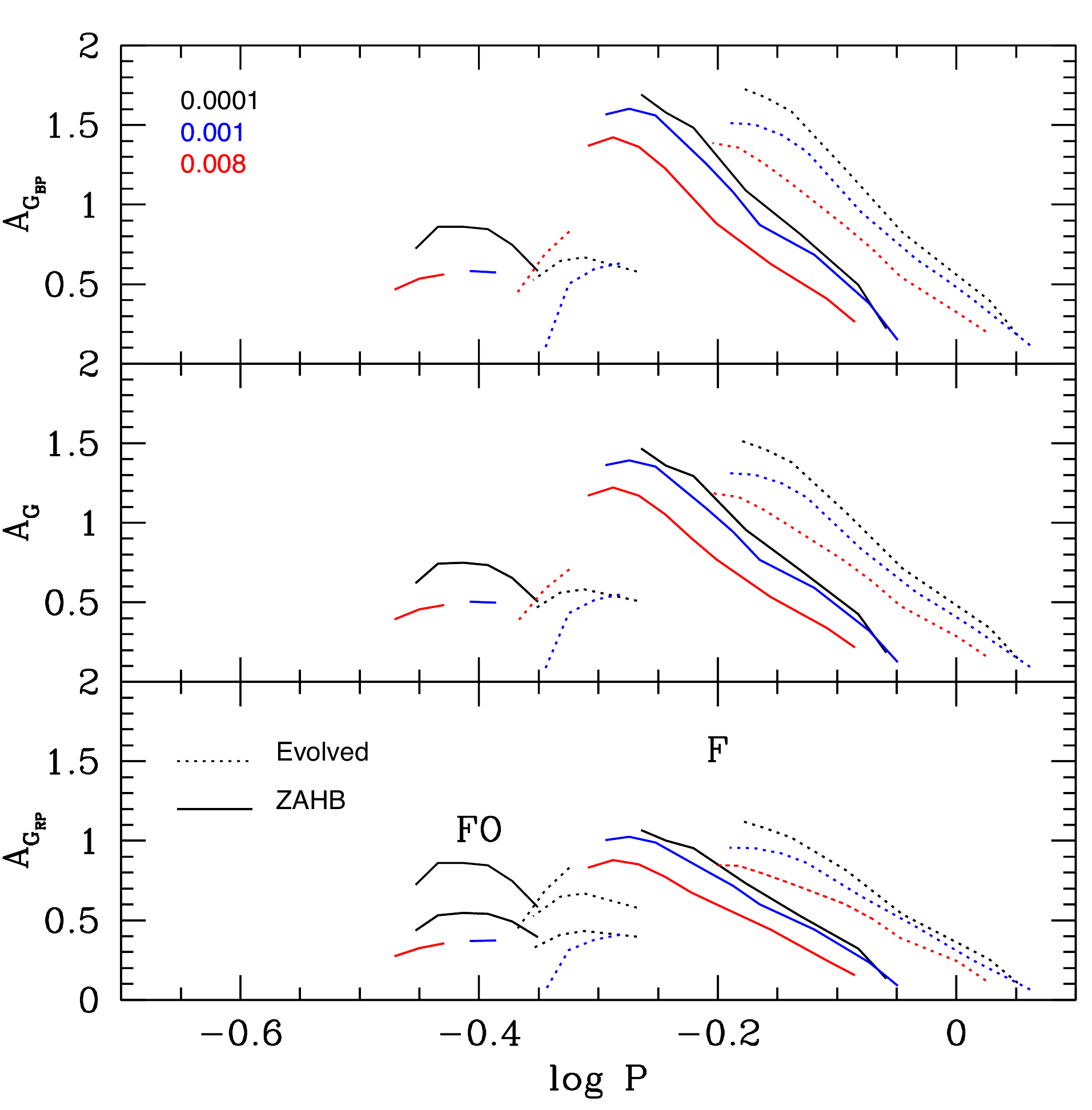}\par 
%\end{multicols}
\caption{\label{fig1} The Bailey diagram in the $G_{BP}$ (top panel), $G$
(middle panel) and $G_{RP}$ (lower panel) filters for the labelled metal abundances and Y=0.30, the corresponding predicted ZAHB masses \citep[see][for details]{marconi15}, namely 0.80$M_{\odot}$ for Z=0.0001,  0.64$M_{\odot}$ for Z=0.001 and 0.56$M_{\odot}$ for Z=0.008, and two luminosity levels corresponding to the ZAHB level (solid line) and a brighter luminosity by 0.1 dex (dotted line), for each fixed mass.}
\end{figure}

\begin{figure}
%\begin{multicols}{3}
    \includegraphics[width=8cm]{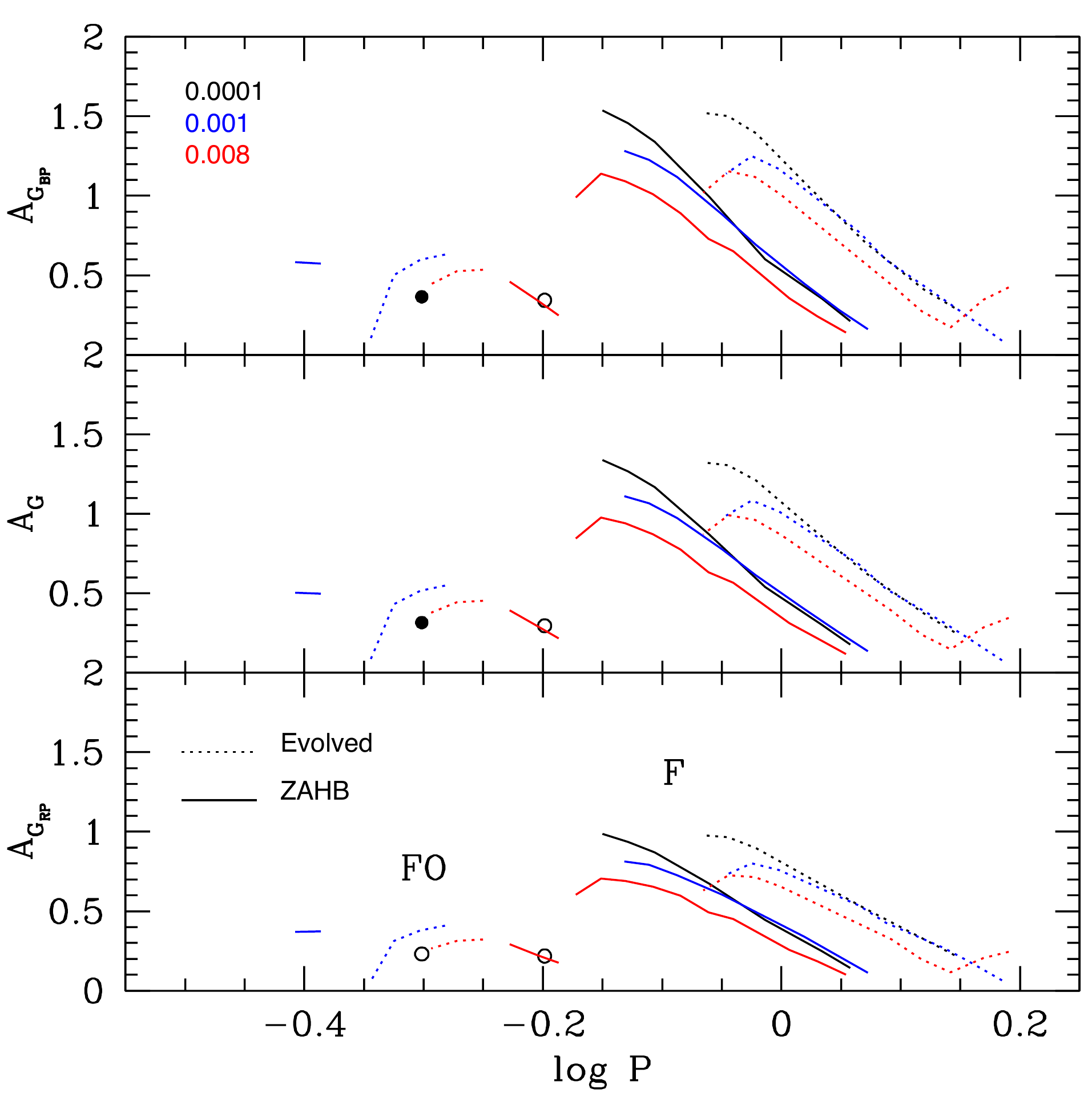}\par 
%\end{multicols}
\caption{\label{fig1} The Bailey diagram in the $G_{BP}$ (top panel), $G$
(middle panel) and $G_{RP}$ (lower panel) filters for the labelled metal abundances and Y=0.40, the corresponding predicted ZAHB masses \citep[see][for details]{marconi15}, namely 0.69$M_{\odot}$ for Z=0.0001,  0.62$M_{\odot}$ for Z=0.001 and 0.55$M_{\odot}$ for Z=0.008, and two luminosity levels corresponding to the ZAHB level (solid line) and a brighter luminosity by 0.1 dex (dotted line), for each fixed mass.
In the case of FO models with $Z=0.0001$ and $Y=0.40$ we find only one case for each of the two luminosity levels, that are indicated as filled (ZAHB) and open (evolved) circles.}
\end{figure}

By including the He-enriched pulsation models computed in \citet{marconi18a}
we can investigate the effect of a possible helium enrichment.
In Figures 11 and 12 we reproduce the Gaia filters Bailey diagram but assuming $Y=0.30$ and $Y=0.40$, respectively.
By comparing these plots with the standard helium case (Figure 9) we notice that, as the helium abundance increases, two main trends occur:
\begin{enumerate}
\item The periods get systematically longer as an effect of the increased ZAHB luminosity level \citep[see e.g.][and references therein]{marconi18a,marconi18b}
\item The pulsation amplitudes get systematically smaller, mainly  as an effect of the reduced hydrogen abundance.
\end{enumerate}

\subsection{Mean magnitudes and colors}

From the theoretical RR Lyrae Gaia filter light curves  we can derive
intensity weighted mean magnitudes. These are reported, for each
individual model, in Tables 1 and 2, for the F and FO-mode, respectively.
The various columns in these tables report the metal and helium abundances, the predicted pulsation period, the input mass, luminosity (in solar units),  effective temperature and the inferred mean magnitudes and pulsation amplitude in the three filters $G_{BP}$, $G$ and $G_{RP}$. On this basis the color index $<G_{BP}>-<G_{RP}> $ can also be derived, for each pulsation model, and the theoretical PW relations can be computed for each individual chemical composition or directly including a metallicity and a helium abundance term, as discussed in the following Section.

\section{The theoretical Period-Wesenheit relations}
From the intensity-weighted mean magnitudes reported in Tables 1 and 2,
we can derive the first theoretical PW relations in the Gaia filters
for RR Lyrae stars, as a function of the metal abundance.  The definition of the adopted Wesenheit relation is the same as in De Somma et al. (2020), namely $G-1.9(<G_{BP}>-<G_{RP}>)$, that in turn was based on the derivation by Ripepi et al. (2019). Additional
relations, including Helium enriched models, and thus providing the
dependence on $Y$ as well, are also derived.
The coefficients of the  PW relations including only the
metallicity term or both the metallicity and the helium abundance,  are reported in Tables 3 and 4, respectively. 
These relations are derived both separately for the two pulsation modes (first two lines of Tables 3 and 4) and globally, by fundamentalizing FO periods \citep[see][and references therein]{marconi15,Coppola15} according to the relation $\log{P_F}=\log{P_{FO}}+0.127$ (last line of Tables 3 and 4).

\onecolumn
%%%%%% 30 June 2020: The complete tables are contained in the files ModelParams.F.tex, ModelParams.FO.tex
\begin{longtable}{cccccccccccc}
  \hline
$Z$ & $Y$ & $P$ & $M$ & $\log(L/L_\odot)$ & $T_e$ & $<BP>$ & $Amp(BP)$ & $<G>$ & $Amp(G)$ & $<RP>$ & $Amp(RP)$ \\ 
 &  & $(days)$ & $(M_\odot)$ & $(dex)$ & $(K)$ & $(mag)$ & $(mag)$ & $(mag)$ & $(mag)$ & $(mag)$ & $(mag)$ \\ 
  \hline
  \endhead
0.0001 & 0.245 & 0.9800 & 0.80 & 1.860 & 5900 & 0.173 & 0.260 & -0.056 & 0.219 & -0.442 & 0.163 \\ 
  0.0001 & 0.245 & 0.9261 & 0.80 & 1.860 & 6000 & 0.162 & 0.527 & -0.059 & 0.452 & -0.434 & 0.343 \\ 
  0.0001 & 0.245 & 0.8776 & 0.80 & 1.860 & 6100 & 0.149 & 0.656 & -0.062 & 0.566 & -0.424 & 0.430 \\ 
  0.0001 & 0.245 & 0.8292 & 0.80 & 1.860 & 6200 & 0.136 & 0.936 & -0.065 & 0.811 & -0.413 & 0.611 \\ 
  0.0001 & 0.245 & 0.7444 & 0.80 & 1.860 & 6400 & 0.108 & 1.157 & -0.070 & 1.014 & -0.383 & 0.783 \\ 
  0.0001 & 0.245 & 0.6692 & 0.80 & 1.860 & 6600 & 0.092 & 1.577 & -0.062 & 1.373 & -0.338 & 1.017 \\
   &  &  & &  & & ... & &  &  & &\\
   \hline
  \caption{Fundamental model parameters: $Z$ and $Y$ values are listed in columns 1 and 2, respectively, the pulsational period of the model is in column 3,  the mass, luminosity and effective temperatures are listed in columns 4,5 and 6, while the Gaia light curve intensity average magnitudes and amplitudes are reported in columns 7, 8, for the $BP$ band, 9, 10 for the $G$ band and 11, 12 for the $RP$ band. The complete table is available in electronic format.}
\end{longtable}

\begin{longtable}{cccccccccccc}
  \hline
$Z$ & $Y$ & $P$ & $M$ & $\log(L/L_\odot)$ & $T_e$ & $<BP>$ & $Amp(BP)$ & $<G>$ & $Amp(G)$ & $<RP>$ & $Amp(RP)$ \\ 
 &  & $(days)$ & $(M_\odot)$ & $(dex)$ & $(K)$ & $(mag)$ & $(mag)$ & $(mag)$ & $(mag)$ & $(mag)$ & $(mag)$ \\ 
  \hline
  \endhead
0.0001 & 0.245 & 0.4720 & 0.80 & 1.860 & 6700 & 0.072 & 0.723 & -0.078 & 0.630 & -0.346 & 0.476 \\ 
  0.0001 & 0.245 & 0.4496 & 0.80 & 1.860 & 6800 & 0.061 & 0.784 & -0.078 & 0.681 & -0.329 & 0.507 \\ 
  0.0001 & 0.245 & 0.4274 & 0.80 & 1.860 & 6900 & 0.052 & 0.853 & -0.077 & 0.736 & -0.312 & 0.531 \\ 
  0.0001 & 0.245 & 0.4082 & 0.80 & 1.860 & 7000 & 0.044 & 0.808 & -0.076 & 0.697 & -0.295 & 0.499 \\ 
  0.0001 & 0.245 & 0.3897 & 0.80 & 1.860 & 7100 & 0.034 & 0.499 & -0.079 & 0.431 & -0.284 & 0.309 \\ 
  0.0001 & 0.245 & 0.4107 & 0.80 & 1.760 & 6600 & 0.339 & 0.686 & 0.178 & 0.594 & -0.109 & 0.441 \\ 
   &  &  & &  & & ... & &  &  & & \\
   \hline
  \caption{First overtone model parameters: $Z$ and $Y$ values are listed in columns 1 and 2, respectively, the pulsational period of the model is in column 3,  the mass, luminosity and effective temperatures are listed in columns 4,5 and 6, while the Gaia light curve intensity average magnitudes and amplitudes are reported in columns 7, 8, for the $BP$ band, 9, 10 for the $G$ band and 11, 12 for the $RP$ band. The complete table is available in electronic format.}
\end{longtable}
\twocolumn

\begin{table*}
\caption{\label{PWZ} The coefficients of the metal-dependent PW relations 
$W = a + b \log {P} +c {\rm [Fe/H]}$ 
for F and FO models. The last column represents the root-mean-square-deviation ($\sigma$) coefficient}
\centering
\begin{tabular}{cccccccc}
\hline\hline
mode&a&b&c&$\sigma_{a}$&$\sigma_{b}$&$\sigma_{c}$&$\sigma$\\
\hline
F &-0.936&-2.296&0.124&0.049&0.032&0.005&0.05\\
FO &-1.344&-2.440&0.112&0.033&0.028&0.005&0.03\\
GLOBAL &-0.952&-2.271&0.123&0.051&0.024&0.004&0.05\\
\hline
\end{tabular}
\end{table*}
 
\begin{table*}
\caption{\label{PWZY} The coefficients of the metal- and helium-dependent PW relations $W = a + b \log{P} +c {\rm [Fe/H]} + d \log{Y}$ for F and FO models}. The last column represents the root-mean-square-deviation ($\sigma$) coefficient.
\centering
\begin{tabular}{cccccccccc}
\hline\hline
mode&a&b&c&d&$\sigma_{a}$&$\sigma_{b}$&$\sigma_{c}$&$\sigma_{d}$&$\sigma$\\
\hline
F &-1.277&-2.298&0.123&-0.573&0.044&0.014&0.003 &0.028 &0.04\\
FO &-1.600&-2.436&0.120&-0.442&0.031&0.020&0.003 &0.031 &0.03\\
GLOBAL &-1.278&-2.257&0.126& -0.558 &0.047&0.012&0.002&0.025& 0.05\\
\hline
\end{tabular}
\end{table*}

In Figure 13  we plot the predicted Gaia band Wesenheit relations, varying the metallicity from $Z=0.0001$ to $Z=0.02$, at standard helium content, for F and FO models (upper
panel)  and combining F with fundamentalized FO models in a global relation (bottom panel).

\begin{figure}
%\begin{multicols}{3}
    \includegraphics[width=8cm]{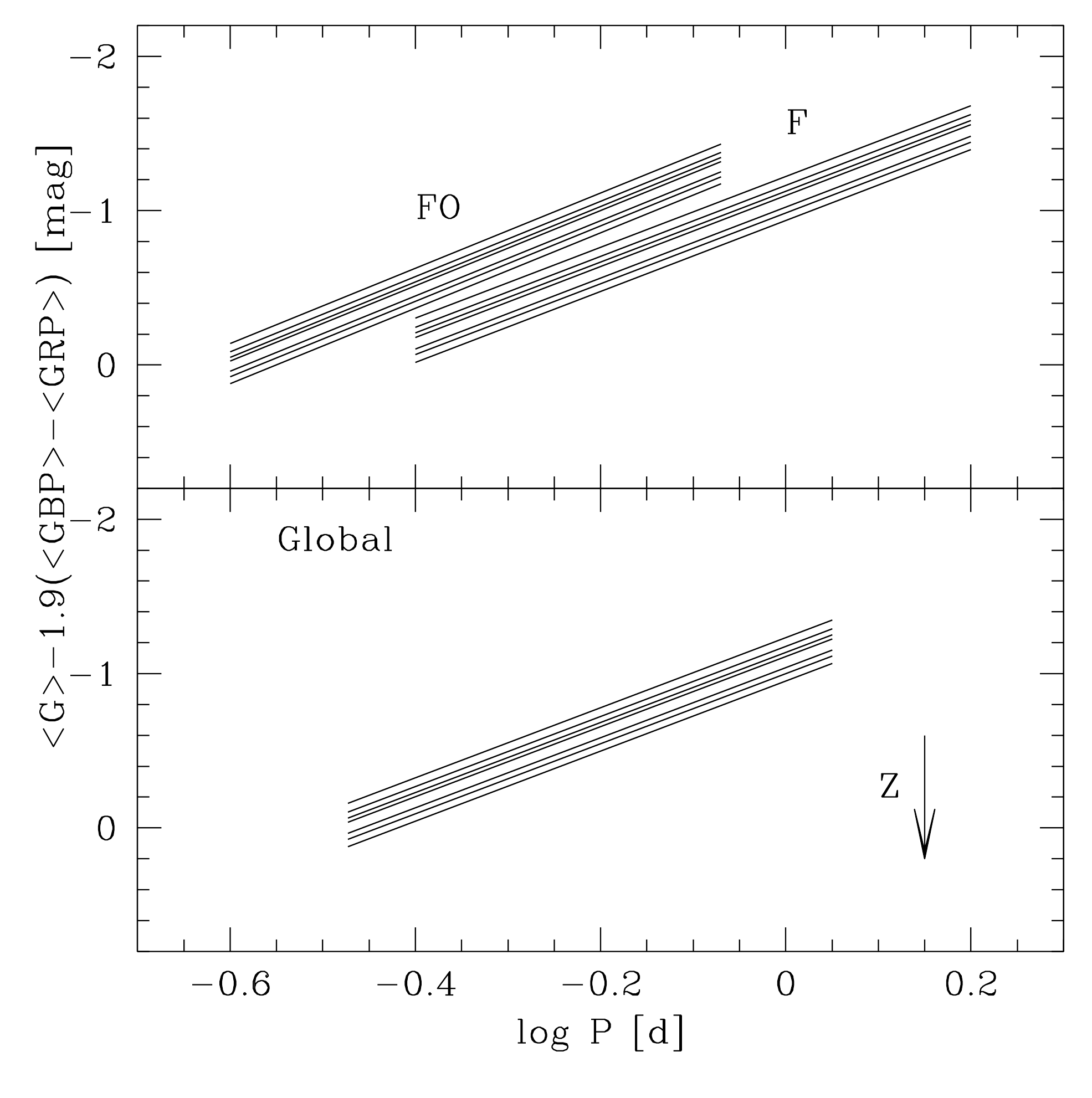}\par 
%\end{multicols}
\caption{\label{fig13} The theoretical PW relations in the Gaia filters, varying the metal abundance from $Z=0.0001$ to $Z=0.02$ (more metallic relations are fainter, see the labelled arrow) and standard Y \citep[as in][]{marconi15}, for F and FO models (upper
panel)  and combining F with fundamentalized FO models in a global relation (bottom panel).}
\end{figure}

We notice that a metallicity variation can change the zero point of the relation by a few tenths of magnitude. In particular, the zero point gets fainter as the metallicity increases (see labelled arrow).

The effect of a variation in the helium content is shown in Figure 14.  Here the same relations presented in Figure 13 are compared with their counterparts for $Y=0.30$ (red lines) and $Y=0.40$ (blue lines), respectively, for the F (top panel) and  FO (middle panel) mode, as well as for the global selection (bottom panel). According to these plots  $Y$ seems to have a minor effect on the slope and the zero point of PW relations even if the period range gets systematically longer and the Wesenheit functions systematically brighter as the helium content increases. These trends are due to the effects of the increased ZAHB luminosity level on the pulsation periods and mean magnitudes as the helium content increases \citep[see][for details]{marconi18a}.
 
\begin{figure}
%\begin{multicols}{3}
    \includegraphics[width=8cm]{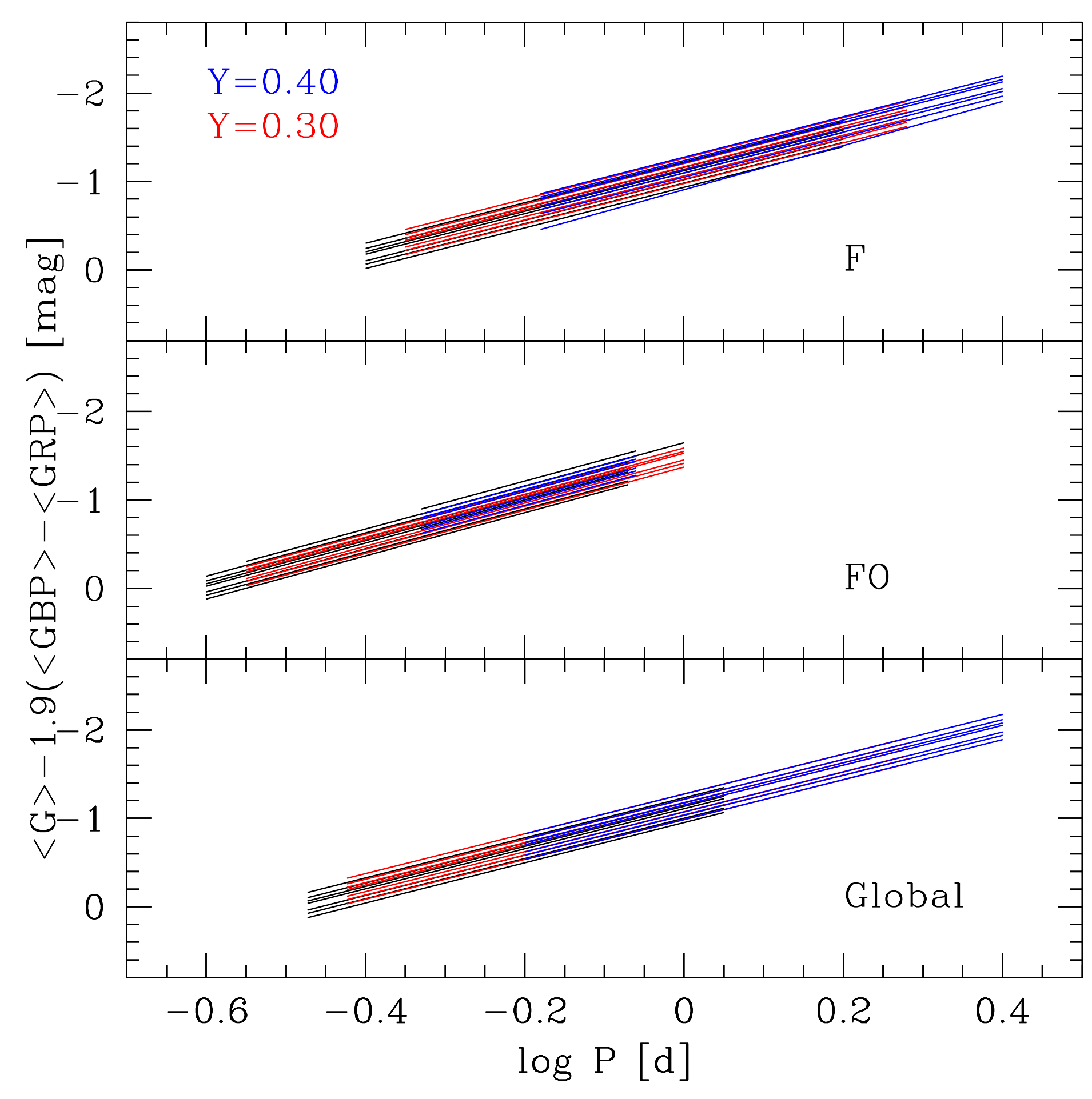}\par 
%\end{multicols}
\caption{\label{fig14} The theoretical PW relations in the Gaia filters, varying the Helium abundance from the standard value (as in Fig. 13, black lines)  to $Y=0.30$ (red lines) and $Y=0.40$ (blue lines) for the F (top panel) and  FO (middle panel) mode, as well as for the global selection (bottom panel).}
\end{figure}

\section{Predicted parallaxes for Gaia RR Lyrae targets}

\subsection{Selection of the sample}

To test our new predictions we searched the literature for RR Lyrae with both a metallicity estimate and {\it Gaia} DR2  photometry in the $G,G_{BP},G_{RP}$ bands calculated as intensity-averaged magnitudes \citep[see][for details]{Holl2018,Clementini2019}. More in detail, we scanned the literature searching for RR Lyrae whose metallicity was estimated on the basis of high-resolution spectroscopy (HRS), in order to guarantee accuracy and precision in the measurement. To this aim we first adopted the compilation by \citet{Magurno2018}, who listed the HRS iron abundances present in the literature for a sample of 134 RR Lyrae stars. An inspection of their Table 10 showed that several objects were observed two or more times by different Authors. In these cases we averaged the results and took the standard deviation as a measure of the uncertainty. As no error is available for the stars with one single measurement, we considered the errors derived above for the pulsators with at least three independent measurements and calculated the mean, obtaining an average error of 0.13 dex. We therefore assigned this value as minimum uncertainty for stars with single measurements, and, to be conservative, extended this uncertainty also to stars with only two measurements whose semi-difference was smaller than 0.13 dex. The results of this exercise is reported in Table~\ref{tab:observationalData}, where we listed the 98 stars in \citet{Magurno2018} sample with {\it Gaia} DR2 intensity-averaged magnitudes. For completeness, we also   reported  the stars without $G_{BP},G_{RP}$ magnitudes. 

The sample by \citet{Magurno2018} was complemented by searching serendipitous RR Lyrae metallicity measurements among recently published spectroscopic surveys based on HRS, namely APOGEE2-DR16 \citep[Apache Point Observatory Galactic Evolution Experiment 2, Data Release 16][]{Ahumada2020} and GALAH Data Release 2 \citep[GALactic Archaeology with HERMES, DR2][]{Buder2018}. As a result of this search, we found 8 and 61 objects in APOGEE and GALAH, respectively. 
As APOGEE observes in the near infrared and has a lower resolution, we searched for possible systematic differences in iron abundance with respect to GALAH\footnote{No meaningful comparison can be made with \citet{Magurno2018}, neither for APOGEE nor for GALAH, as the overlap is restricted to a couple of stars}. To this aim we cross-correlated the entire catalogues of APOGEE and GALAH, using the resulting 515 stars in common, covering an interval $-$0.5$<${\rm [Fe/H]}$<$0.5 dex, and calculated the following equation to apply a small correction to APOGEE results: ${\rm [Fe/H]}_G=(-0.0155 \pm 0.0044)+ (1.047 \pm 0.016){\rm [Fe/H]}_A$, with rms=0.058 dex, where ${\rm [Fe/H]}_G$ and ${\rm [Fe/H]}_A$ are the iron abundances in the GALAH and APOGEE system, respectively. The iron abundance data with the relative uncertainties for the APOGEE and GALAH survey are shown in Table~\ref{tab:observationalData}. Note that for the APOGEE data, the table lists the corrected [Fe/H] values  and the original uncertainties have been summed in quadrature with the rms error of the relation converting APOGEE into GALAH iron abundanced. The total HRS sample in this table amounts to 167 objects. We applied a further selection to this sample  removing all the objects with negative parallax and with RUWE$\geq$1.4 as suggested by the {\it Gaia} documentation\footnote{The RUWE parameter measures the reliability of the {\it Gaia} astrometry, see Section 14.1.2 of "Gaia Data Release 2 Documentation release 1.2"; https://gea.esac.esa.int/archive/documentation/GDR2/.}. Moreover we selected only RR Lyrae with relative parallax error lower than 10$\%$\footnote{Some authors \citep[see e.g.]{BJ18} assert that when Gaia parallaxes are precise at level of 10\%, they can be used to derive reliable distances.}. Therefore the final HRS sample comprises 103 objects.  

In addition to the above described sample, upon Referee suggestion, we also adopted the sample by \citet{muraveva18}, largely based on the work by \citet{Dambis2013}, that collected literature metallicities for RR Lyrae derived with different techniques from spectroscopic data at distinct resolutions. Applying the same cuts quoted above, the \citet{muraveva18} sample shrinks to 112 objects. 

\begin{center}
\begin{figure}
%\begin{multicols}{3}
\includegraphics[height=8cm]{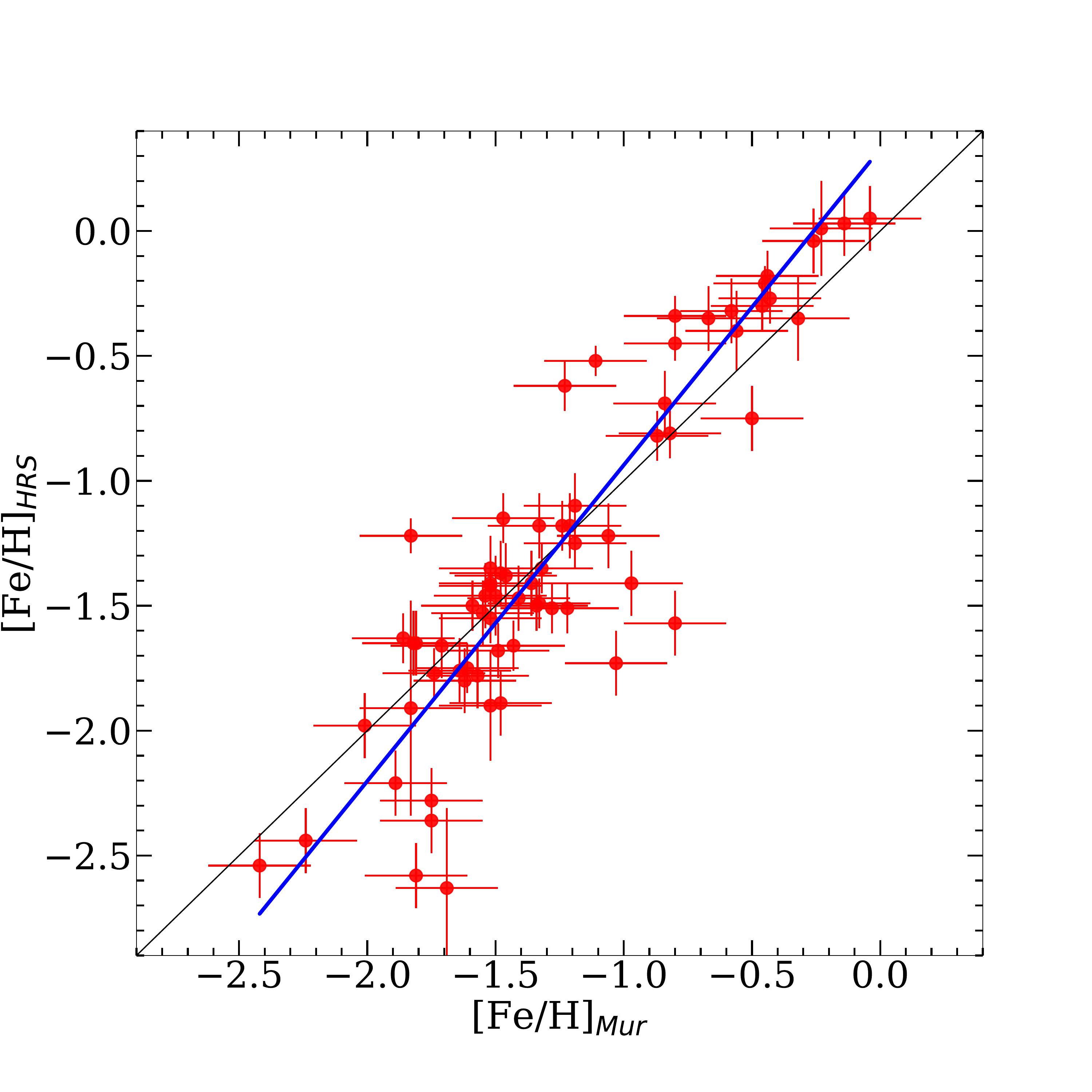}\par 
%\end{multicols}
\caption{\label{fig:comparison} Comparison between the HRS and \citet{muraveva18} samples. The 45\degr is shown a thin black line. The linear regression is displayed as a thick blue line.}
\end{figure}
\end{center}

Before proceeding, it is worth comparing the HRS and \citet{muraveva18} samples. 
There are 70 stars in common between these samples. The correlation between the iron abundances is shown in Fig.~\ref{fig:comparison}. It can be easily seen that there is a detectable trend with metallicity between the two samples. The linear correlation between the iron content values in the two samples is: ${\rm [Fe/H]}_{HRS}=(0.328 \pm 0.092)+ (1.265 \pm 0.069){\rm [Fe/H]}_{Mur}$, with rms=0.21 dex, where ${\rm [Fe/H]}_{HRS}$ and ${\rm [Fe/H]}_{Mur}$ are the iron abundances in the HRS and \citet{muraveva18} samples, respectively.

\subsection{Application of theoretical PWZ relations}

In this section we apply to the selected HRS and \citet{muraveva18} samples  the theoretical PWZ relations reported in Table 3 for the F, FO and global case. The application of the quoted PWZ relations allows us to derive individual distance moduli and, in turn, individual parallaxes to be compared with Gaia DR2 determinations. The differences between the theoretical parallaxes and Gaia DR2 results are shown in Figures 16 and 17, for the HRS and \citet{muraveva18} samples, respectively, when applying the F (upper panel), FO (middle panel) and global (bottom panel) PWZ relations. In both figures, blue and red symbols correspond to accepted and discarded objects by a 2.5 $\sigma-$clipping procedure. 
The error bars take into account the observational parallax error and the intrinsic dispersion of the adopted PWZ relations. The labelled mean weighted differences suggest a very good agreement between theoretical and empirical distance determinations. 
These results are consistent with the zero-point offset obtained for RR Lyrae by \citet{Arenou2018}, who validated Gaia DR2 catalogue finding a negligible ($-$0.01$\pm$0.02 mas) 
offset between the HST and DR2
parallaxes, and slightly smaller than the offset obtained by \citet{muraveva18} (-0.057 mas) for RR Lyrae, by \citet{Riess2018} ($-0.046\pm$0.013 mas) and \citet{Ripepi2019} ($-0.07$ mas) for Classical Cepheids. and by \citet{desomma2020} from the application of theoretical PW relations at solar chemical composition to a sample of Gaia DR2 Galactic Cepheids, even if still consistent within the  errors.
Indeed, a different zero-point offset might, in principle, be obtained for Cepheids and RR Lyrae as an effect of its possible dependence on magnitude and color but new more accurate parallaxes, as expected from Gaia Data Release 3 and/ subsequent releases, are needed in order to properly fix this quantity.

\begin{figure}
%\begin{multicols}{3}
    \includegraphics[width=8cm]{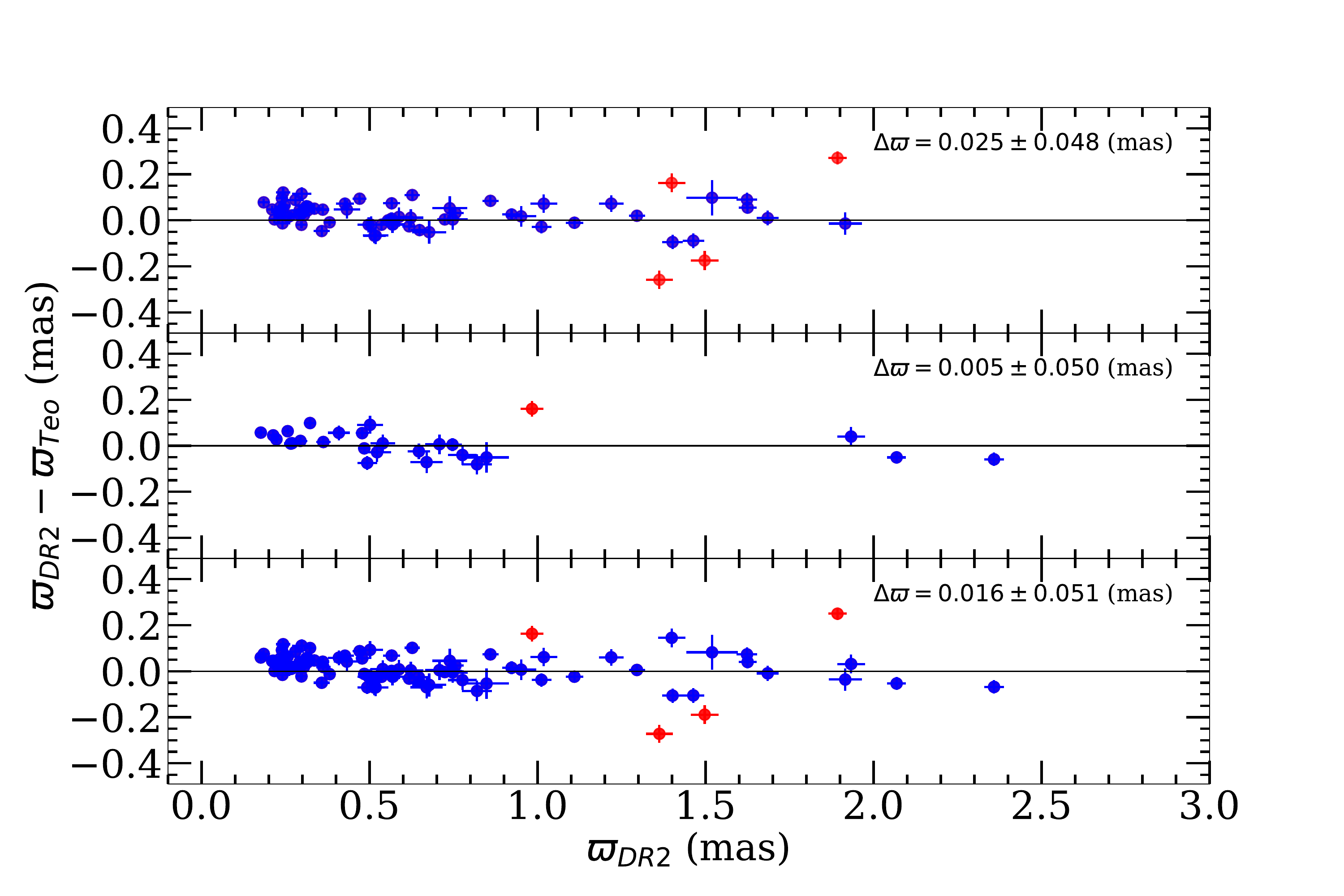}\par 
%\end{multicols}
\caption{\label{fig15} Difference between the theoretical parallaxes and Gaia DR2 results for the F (upper panel), FO (middle panel) and global sample selections,  respectively, taking into account the HRS sample discussed in the text. Blue and red symbols correspond to accepted and discarded objects according to a 2.5$\sigma$-clipping procedure.}
\end{figure}

\begin{figure}
%\begin{multicols}{3}
    \includegraphics[width=8cm]{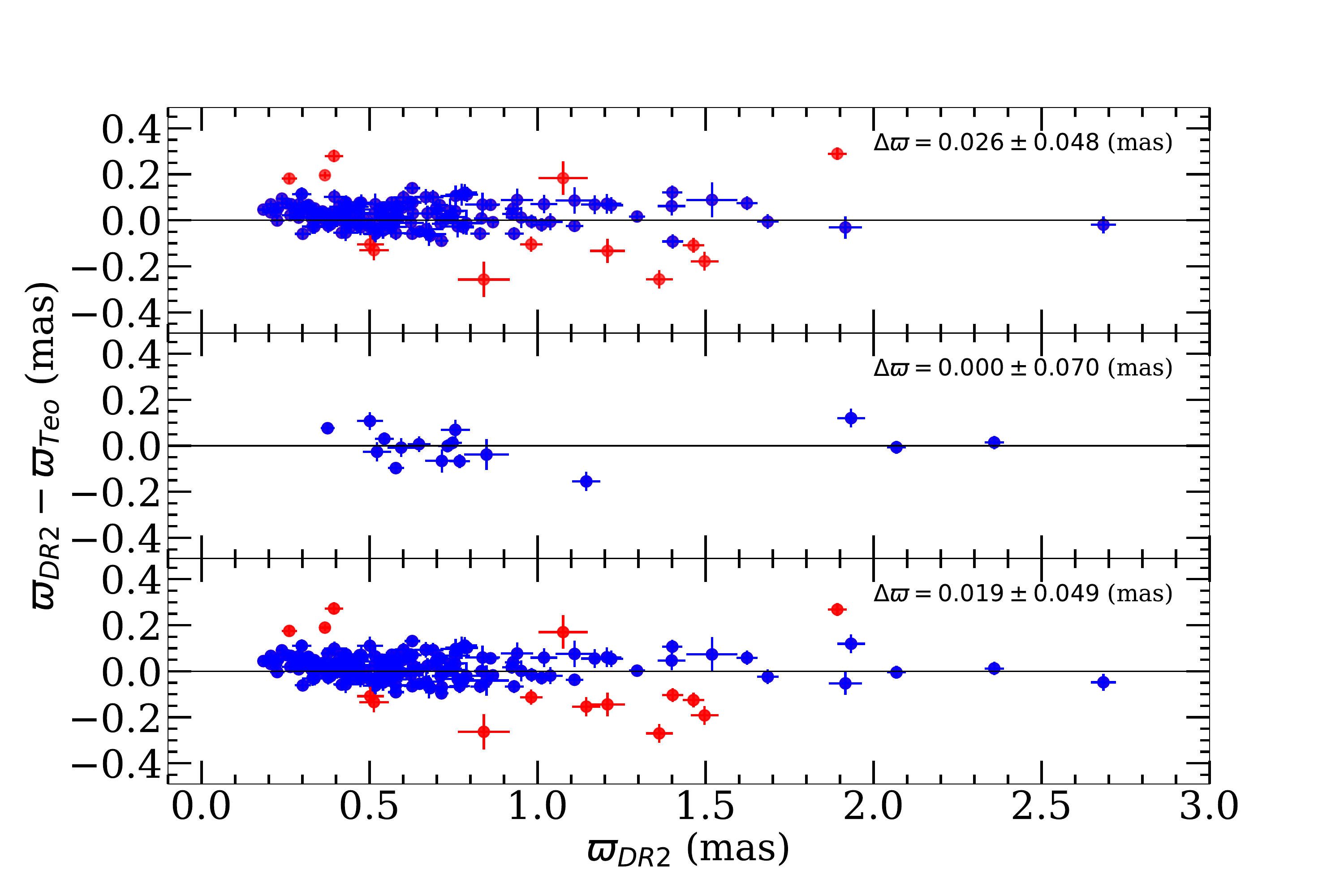}\par 
%\end{multicols}
\caption{\label{fig15} Difference between the theoretical parallaxes and Gaia DR2 results for the F (upper panel), FO (middle panel) and global sample selections, respectively, taking into account the \citet{muraveva18} sample discussed in the text. Blue and red symbols correspond to accepted and discarded objects according to a 2.5$\sigma$-clipping procedure.}
\end{figure}

%\section{An inferred metallicity distribution of Gaia Galactic RR Lyrae}
%
%In this Section we assume Gaia DR2 parallaxes for the Galactic RR Lyrae of %the investigated sample and invert the theoretical PWZ relations reported %in Table 3 to provide independent individual metal abundances.
%The obtained {\rm [Fe/H]} istogram for F and FO pulsators are shown in Figure 16 %(upper panel) as compared with the metallicity distribution by XXXX.
%The lower panel of the same figure shows the inferred metallicity %distribution when the global PWZ relation is inverted for F and %fundamentalized FO RR Lyrae.
%XXXX TBC XXXX

\section{Conclusions}
An extensive set of nonlinear convective pulsation models for RR Lyrae at different metal and helium abundances has  been taken into account. The transformation of bolometric magnitude variations into  the Gaia filters allowed us to derive the first theoretical light curves directly comparable with Gaia time-series data. In particular, we built the first theoretical Bailey diagrams  and PW relations in the $G_{BP}$, $G$ and $G_{RP}$ filters, varying both the metallicity and the helium content.
As for the Bailey diagram we conclude that an increase in the metal abundance and/or in the helium abundance produces a decrease in the pulsation amplitudes, whereas an increase in the luminosity level produces a period shift towards longer values. In particular, in the case of FO RR Lyrae, the location of the described bell-shape in the Bailey diagram can be used to constrain the luminosity level \citep[see e.g.][]{BCCM1997}. 
The theoretical PW relations in the Gaia bands show a dependence of the zero point on metal abundance, in the sense that brighter Wesenheit functions are predicted for more metal poor chemical composition and a lower effect due to variations in the helium content, with helium enriched models characterized by longer periods and brighter Wesenheit functions.
The theoretical PWZ relations are applied to a subset of Gaia DR2 RR Lyrae (293 F and 50 FO pulsators) with complementary metallicity information   to infer individual theoretical parallaxes, that are in very good agreement with Gaia results. In particular, the inferred zero-point parallax offset is consistent with zero both in the case of F and FO pulsators. 
Even if more stringent conclusions could be drawn in the future from the next Gaia data releases, the obtained results seem on one side to support the predictive capabilities of current pulsation models and on the other to suggest that a smaller parallax offset could be required for the bluer older and lower mass RR Lyrae stars than for Classical Cepheids.

\section*{Acknowledgements}

   We thank an anonymous referee for her/his useful comments. This    work    has    made    use    of    data from  the  European  Space  Agency  (ESA)  mission Gaia (https://www.cosmos.esa.int/gaia),   processed  by  the Gaia Data  Processing  and  Analysis  Consortium  (DPAC,  https://www.cosmos.esa.int/web/gaia/dpac/consortium).   Funding for  the  DPAC  has  been  provided  by  national  institutions, in  particular  the  institutions  participating  in  the  Gaia  Multilateral  Agreement.   In  particular,  the  Italian  participation in  DPAC  has  been  supported  by  Istituto  Nazionale  di  Astrofisica  (INAF)  and  the  Agenzia  Spaziale  Italiana  (ASI) through grants I/037/08/0, I/058/10/0, 2014-025-R.0, 2014-025-R.1.2015  and  2018-24-HH.0  to  INAF  (PI  M.G.  Lattanzi).  We acknowledge partial financial support from ’Progetto Premiale’ MIUR MITIC (PI B. Garilli) and the INAF Main Stream SSH program,  1.05.01.86.28. We   acknowledge   Istituto   Nazionale   di Fisica   Nucleare   (INFN),   Naples   section,    specific   initiative    QGSKY.   This work has made  use  of  the  VizieR  database,  operated  at  CDS,  Strasbourg, France

\section{Data Availability}
The multi-band  model light curves data are available upon request to the authors. Tables 1 and 2 are published in electronic form.

%%%%%%%%%%%%%%%%%%%%%%%%%%%%%%%%%%%%%%%%%%%%%%%%%%



\bibliographystyle{mnras}


%%%%%%%%%%%%%%%%%%%%%%%%%%%%%%%%%%%%%%%%%%%%%%%%%%

%%%%%%%%%%%%%%%%% APPENDICES %%%%%%%%%%%%%%%%%%%%%

\appendix

\section{Sample of RR Lyrae with metallicity from high resolution spectroscopy.}

%In this appendix we discuss the homogeneization of the metallicity scales 
%between the different surveys adopted to build up the observational sample. 
%First we choose the GALAH survey as reference because it is the work with the highest resolution and likely the highest accuracy. Secondly, we downloaded the whole releases of each of the adopted surveys and cross-matched all the possible stars with the GALAH to obtain a data-set as large as possible to derive the transformation between the different metallicity systems. Third, we carried out linear fits to correlate the metallicity of each survey vs GALAH. Finally, we used the derived relationships to put all the metallicities in a common system based on GALAH. To obtain these relations we adopted the {\sc python} code {\sc LtsFit}\footnote{https://pypi.org/project/ltsfit/} \citep{Cappellari2013} which performs robust fits and allows to take into account errors on both axis. The derived relations are summarized in Table~\ref{tab:linearFits} where we also provide the number of stars used, the metallicity range and the pivoting metallicity, if needed (see {\sc LtsFit} documentation).    

%\begin{landscape}
\begin{table*}
\caption{Sample of RR Lyrae stars used for the comparison with the models with metallicity from high resolution spectroscopy. The meaning of the columns is: (1) {\it Gaia} source identifiers; (2) mode of pulsation: RRab, RRc and RRd mean fundamental, first overtone and double mode pulsator, respectively; (3,4) coordinates; (5,6) {\it Gaia} parallax and relative error; (7) ruwe parameter (see text); (8) Period; (9,10,11) $G,G_{BP},G_{RP}$ intensity-averaged magnitudes; (12) number of averaged measures: for APOGEE and GALAH survey it was imposed equal to one; (13,14) adopted iron abundance and relative error; (15) source of the iron abundance: APO=APOGEE; GAL=GALAH; M18=\citet{Magurno2018}.}     
\label{tab:observationalData}
%\centering
\setlength{\tabcolsep}{2.5pt}
\begin{tabular}{llrrccccccccrcc}
\hline\hline
~~~~~~~~~~~~~~~~~~ID    &    Mode  &     RA    &    Dec   &  $\varpi $ & $\sigma_\varpi$  & ruwe   &  P    &       G  &       GBP   &   GRP &  n & [Fe/H]& $\sigma$[Fe/H] & Source\\
    &          &    (J2000)  & (J2000)   &   (mas) &  (mas)  &        & days  &     (mag)&      (mag)  &  (mag) &  & (dex) &   (dex)     &         \\
~~~~~~~~~~~~~~~~~~(1)    &~~(2)    & (3)    &      (4)    & 	   (5)    &    (6)    &   (7)    &   (8)    & 	  (9)    &    (10)    &  (11)    &   (12)    & (13)    & (14)   & (15) \\	      
\hline
  ASAS\_J184654-5439.3          &    RRc   &  281.72510  &  $-$54.65558  &  0.3288  &  0.0348  &  1.170  &   0.23413   &   13.303   &   13.431   &   13.038  &  1  &  $-$0.46   &   0.09    &  GAL  \\
  ASAS\_J164128-1029.6          &    RRc   &  250.36483  &  $-$10.49337  &  0.7079  &  0.0428  &  1.039  &   0.23673   &   12.589   &   12.951   &   12.048  &  1  &  $-$0.35   &   0.07    &  GAL  \\
  ASAS\_J202817-3806.6          &    RRc   &  307.07122  &  $-$38.11025  &  0.2621  &  0.0425  &  1.219  &   0.24485   &   13.261   &   13.385   &   13.006  &  1  &  $-$0.47   &   0.07    &  GAL  \\
               KIC8832417      &    RRc   &  296.72630  &   45.08063  &  0.4842  &  0.0188  &  0.936  &   0.24855   &   12.998   &   13.252   &   12.583  &  1  &   $-$0.27  &    0.13   &  M18  \\
               KIC5520878      &    RRc   &  287.59821  &   40.76791  &  0.2227  &  0.0166  &  1.128  &   0.26917   &   14.038   &   14.222   &   13.692  &  1  &   $-$0.18  &    0.13   &  M18  \\
                    YZCap      &    RRc   &  319.88499  &  $-$15.11706  &  0.8480  &  0.0668  &  1.331  &   0.27345   &   11.219   &   11.384   &   10.928  &  2  &   $-$1.51  &    0.10   &  M18  \\
  ASAS\_J023319-7336.7          &    RRc   &   38.32818  &  $-$73.61193  &  0.4779  &  0.0189  &  1.199  &   0.28714   &   11.983   &   12.116   &   11.735  &  1  &  $-$0.77   &   0.07    &  GAL  \\
  GDR2\_5202509386185816704     &    RRc   &  149.90233  &  $-$78.73859  &  0.2562  &  0.0124  &  1.089  &   0.29035   &   13.429   &   13.631   &   13.069  &  1  &  $-$0.74   &   0.08    &  GAL  \\
                     UCom      &    RRc   &  190.01308  &   27.49887  &  0.5223  &  0.0421  &  1.138  &   0.29274   &   11.670   &   11.537   &   11.358  &  1  &   $-$1.41  &    0.13   &  M18  \\
  ASASSN\_J185318.40-542921.7   &    RRc   &  283.32665  &  $-$54.48947  &  0.2691  &  0.0239  &  1.134  &   0.29274   &   13.526   &   13.678   &   13.226  &  1  &  $-$0.48   &   0.09    &  GAL  \\
  CRTS\_J102704.8-412916        &    RRc   &  156.77017  &  $-$41.48780  &  0.2655  &  0.0203  &  1.144  &   0.29278   &   13.576   &   13.737   &   13.266  &  1  &  $-$0.72   &   0.08    &  GAL  \\
       ASAS\_J110522-2641.0     &    RRc   &  166.34138  &  $-$26.68466  &  0.5391  &  0.0371  &  1.078  &   0.29446   &   11.802   &   11.941   &   11.551  &  2  &   $-$1.69  &    0.13   &  M18  \\
  CRTS\_J201409.3-464306        &    RRc   &  303.53934  &  $-$46.71813  &  0.2076  &  0.0325  &  1.206  &   0.29982   &   13.106   &   13.218   &   12.854  &  1  &  $-$0.58   &   0.08    &  GAL  \\
       ASAS\_J200431-5352.3     &    RRc   &  301.13103  &  $-$53.87190  &  0.8197  &  0.0444  &  1.114  &   0.30023   &   11.029   &   11.166   &   10.775  &  2  &   $-$2.69  &    0.13   &  M18  \\
                    RZCep      &    RRc   &  339.80584  &   64.85932  &  2.3585  &  0.0290  &  0.950  &   0.30871   &    9.256   &    9.556   &    8.805  &  1  &   $-$2.36  &    0.13   &  M18  \\
  OGLE-SMC-RRLYR-6212          &    RRc   &   31.91403  &  $-$77.81319  &  0.3228  &  0.0184  &  1.159  &   0.31067   &   12.581   &   12.742   &   12.279  &  1  &  $-$0.62   &   0.08    &  GAL  \\
       ASAS\_J203145-2158.7     &    RRc   &  307.93716  &  $-$21.97964  &  0.7762  &  0.0428  &  1.000  &   0.31071   &   11.320   &   11.494   &   11.025  &  1  &   $-$1.17  &    0.13   &  M18  \\
                    CSEri      &    RRc   &   39.27456  &  $-$42.96331  &  2.0684  &  0.0279  &  1.065  &   0.31133   &    8.924   &    9.080   &    8.676  &  1  &   $-$1.89  &    0.13   &  M18  \\
                    TVBoo      &    RRc   &  214.15240  &   42.35977  &  0.7468  &  0.0285  &  1.032  &   0.31256   &   10.911   &   11.041   &   10.685  &  1  &   $-$2.44  &    0.13   &  M18  \\
  ASAS\_J145747-3812.2          &    RRc   &  224.44490  &  $-$38.20380  &  0.1249  &  0.1364  &  1.154  &   0.31664   &   13.067   &   13.301   &   12.794  &  1  &  $-$0.76   &   0.08    &  GAL  \\
  GDR2\_6367478755093421056     &    RRc   &  291.66128  &  $-$74.64357  &  0.4086  &  0.0321  &  1.204  &   0.31673   &   12.424   &   12.612   &   12.104  &  1  &  $-$0.70   &   0.07    &  GAL  \\
                    MTTel      &    RRc   &  285.55030  &  $-$46.65386  &  1.9331  &  0.0414  &  0.927  &   0.31690   &    8.911   &    9.096   &    8.674  &  1  &   $-$2.58  &    0.13   &  M18  \\
  ASASSN\_J190454.75-643958.7   &    RRc   &  286.22809  &  $-$64.66630  &  0.1288  &  0.0184  &  1.097  &   0.31752   &   13.937   &   14.104   &   13.625  &  1  &  $-$0.73   &   0.09    &  GAL  \\
                     TSex      &    RRc   &  148.36825  &    2.05722  &  1.2466  &  0.0444  &  0.996  &   0.32468   &    9.964   &   99.999   &   99.999  &  2  &   $-$1.66  &    0.10   &  M18  \\
  OGLE-SMC-RRLYR-2621          &    RRc   &  357.91003  &  $-$73.57293  &  0.3623  &  0.0211  &  1.059  &   0.32635   &   12.698   &   12.853   &   12.409  &  1  &  $-$0.87   &   0.08    &  GAL  \\
  ASAS\_J180809-6527.4          &    RRc   &  272.03547  &  $-$65.45624  &  0.2941  &  0.0205  &  1.068  &   0.32735   &   13.241   &   13.424   &   12.907  &  1  &  $-$0.78   &   0.08    &  GAL  \\
                     YCrv      &    RRc   &  189.54342  &  $-$15.00004  &  0.6695  &  0.0478  &  1.208  &   0.32903   &   11.554   &   11.696   &   11.296  &  1  &   $-$1.39  &    0.13   &  M18  \\
  GDR2\_1317846466364172800     &    RRc   &  244.85760  &   29.71312  &  0.6466  &  0.0335  &  1.229  &   0.33169   &   11.328   &   11.469   &   11.065  &  1  &  $-$2.63   &   0.32    &  APO  \\
  GDR2\_6109120799902812928     &    RRc   &  207.67380  &  $-$42.24301  &  0.4929  &  0.0295  &  0.946  &   0.33264   &   12.798   &   13.012   &   12.462  &  1  &   0.33   &   0.07    &  GAL  \\
  GDR2\_1686384274158206208     &    RRc   &  195.87937  &   71.11218  &  0.9834  &  0.0343  &  1.295  &   0.33298   &   10.184   &   10.328   &    9.888  &  1  &  $-$1.62   &   0.18    &  APO  \\
               KIC4064484      &    RRc   &  293.43948  &   39.12052  &  0.1688  &  0.0201  &  1.102  &   0.33700   &   14.421   &   14.642   &   14.024  &  1  &   $-$1.58  &    0.13   &  M18  \\
  ASASSN\_J211212.83-501250.1   &    RRc   &  318.05352  &  $-$50.21394  &  0.2539  &  0.0327  &  1.167  &   0.34175   &   13.806   &   13.946   &   13.519  &  1  &  $-$0.73   &   0.08    &  GAL  \\
                    AAAql      &    RRab  &  309.56279  &   $-$2.89034  &  0.6774  &  0.0506  &  1.045  &   0.36177   &   11.816   &   11.867   &   11.467  &  1  &   $-$0.32  &    0.13   &  M18  \\
               KIC9453114      &    RRc   &  285.96047  &   46.02887  &  0.2132  &  0.0182  &  1.322  &   0.36573   &   13.285   &   13.438   &   12.984  &  1  &   $-$2.13  &    0.13   &  M18  \\
                    SVScl      &    RRc   &   26.24859  &  $-$30.05943  &  0.5014  &  0.0391  &  1.108  &   0.37736   &   11.304   &   11.445   &   11.059  &  1  &   $-$2.28  &    0.13   &  M18  \\
                    RSBoo      &    RRab  &  218.38841  &   31.75461  &  1.3624  &  0.0396  &  1.138  &   0.37737   &   10.331   &   10.369   &   10.078  &  4  &   $-$0.35  &    0.17   &  M18  \\
  GDR2\_5779689627114584576     &    RRc   &  229.25498  &  $-$77.93312  &  0.1762  &  0.0171  &  1.074  &   0.37751   &   13.812   &   14.009   &   13.466  &  1  &  $-$0.61   &   0.09    &  GAL  \\
                    AVPeg      &    RRab  &  328.01170  &   22.57479  &  1.4635  &  0.0320  &  1.136  &   0.39037   &   10.472   &   10.692   &   10.069  &  3  &   $-$0.18  &    0.10   &  M18  \\
  OGLE-SMC-RRLYR-2594          &    RRc   &  356.81960  &  $-$78.70534  &  0.1629  &  0.0177  &  1.122  &   0.39459   &   13.910   &   14.110   &   13.554  &  1  &  $-$0.87   &   0.09    &  GAL  \\
                  V445Oph      &    RRc   &  246.17172  &   $-$6.54162  &  1.6154  &  0.0526  &  1.220  &   0.39703   &   10.837   &   99.999   &   99.999  &  5  &    0.01  &    0.19   &  M18  \\
  ASAS\_J114710-4131.7          &    RRab  &  176.79419  &  $-$41.52862  &  0.3606  &  0.0200  &  1.062  &   0.39741   &   13.157   &   13.409   &   12.745  &  1  &  $-$0.34   &   0.08    &  GAL  \\
                    TWHer      &    RRab  &  268.63002  &   30.41046  &  0.8596  &  0.0238  &  1.151  &   0.39960   &   11.208   &   11.527   &   10.923  &  1  &   $-$0.35  &    0.13   &  M18  \\
  GDR2\_5821920567383108224     &    RRab  &  244.04851  &  $-$65.81513  &  0.3047  &  0.0180  &  1.190  &   0.40542   &   13.676   &   13.936   &   13.265  &  1  &   0.08   &   0.08    &  GAL  \\
                    CNLyr      &    RRab  &  280.31643  &   28.72253  &  1.1096  &  0.0262  &  1.050  &   0.41138   &   11.260   &   11.584   &   10.784  &  1  &   $-$0.04  &    0.13   &  M18  \\
                     WCrt      &    RRab  &  171.62345  &  $-$17.91441  &  0.7475  &  0.0449  &  1.292  &   0.41201   &   11.450   &   11.650   &   11.138  &  1  &   $-$0.75  &    0.13   &  M18  \\
  ASASSN\_J154554.85-401900.1   &    RRab  &  236.47853  &  $-$40.31668  &  0.2106  &  0.0332  &  1.056  &   0.41335   &   14.309   &   14.815   &   13.779  &  1  &   0.29   &   0.09    &  GAL  \\
  ASAS\_J005001-6238.1          &    RRd   &   12.50263  &  $-$62.63541  &  0.6271  &  0.0247  &  1.216  &   0.41453   &   12.088   &   12.383   &   11.847  &  1  &  $-$0.45   &   0.07    &  GAL  \\
  ASAS\_J045314-3749.2          &    RRab  &   73.31013  &  $-$37.82105  &  0.4704  &  0.0209  &  1.228  &   0.41978   &   12.043   &   12.369   &   11.877  &  1  &  $-$0.52   &   0.06    &  GAL  \\
                    DMCyg      &    RRab  &  320.29810  &   32.19129  &  0.9652  &  0.0507  &  1.479  &   0.41987   &   11.439   &   11.783   &   11.086  &  1  &    0.03  &    0.13   &  M18  \\
                    ARPer      &    RRab  &   64.32163  &   47.40014  &  1.9156  &  0.0491  &  1.005  &   0.42556   &   10.237   &   10.615   &    9.694  &  3  &   $-$0.27  &    0.10   &  M18  \\
                  V440Sgr      &    RRab  &  293.08657  &  $-$23.85378  &  1.3991  &  0.0408  &  0.993  &   0.42893   &   10.158   &   10.638   &    9.922  &  2  &   $-$1.15  &    0.10   &  M18  \\
  GDR2\_6362257964645972352     &    RRab  &  305.14572  &  $-$78.30634  &  0.3064  &  0.0157  &  1.201  &   0.43072   &   13.449   &   13.729   &   12.998  &  1  &  $-$0.62   &   0.09    &  GAL  \\
                  V839Cyg      &    RRab  &  290.07867  &   47.13012  &  0.2171  &  0.0174  &  1.065  &   0.43378   &   14.454   &   14.727   &   14.052  &  1  &   $-$0.05  &    0.13   &  M18  \\
                 V1104Cyg      &    RRab  &  289.50206  &   50.75495  &  0.1024  &  0.0237  &  1.216  &   0.43639   &   14.635   &   14.837   &   14.307  &  1  &   $-$1.23  &    0.13   &  M18  \\
  CRTS\_J171304.1+355841        &    RRab  &  258.26658  &   35.97854  &  0.6270  &  0.0226  &  1.058  &   0.44036   &   11.380   &   11.767   &   11.215  &  1  &  $-$1.73   &   0.13    &  APO  \\
                    KXLyr      &    RRd   &  278.31341  &   40.17304  &  0.9268  &  0.0239  &  1.032  &   0.44091   &   10.889   &   11.132   &   10.598  &  2  &   $-$0.30  &    0.10   &  M18  \\
  CSS\_J165135.0-040010         &    RRab  &  252.89633  &   $-$4.00300  &  0.2429  &  0.0317  &  1.081  &   0.44571   &   14.279   &   14.718   &   13.920  &  1  &  $-$0.52   &   0.09    &  GAL  \\
                    VXHer      &    RRab  &  247.66978  &   18.36691  &  0.9797  &  0.0589  &  1.226  &   0.45536   &   10.774   &   99.999   &   99.999  &  5  &   $-$1.42  &    0.13   &  M18  \\
                    RVUMa      &    RRab  &  203.32515  &   53.98720  &  0.9227  &  0.0277  &  1.116  &   0.46806   &   10.693   &   10.891   &   10.412  &  2  &   $-$1.25  &    0.10   &  M18  \\
\hline
\end{tabular}
\end{table*}
%\end{landscape}

\begin{table*}
\contcaption{}
%\caption{Sample of RR Lyrae stars used for the comparison with the models with metallicity from high resolution spectroscopy additional with respect to \citet{Magurno2018}. The meaning of the columns is: (1) {\it Gaia} source identifiers; (2) mode of pulsation: RRab, RRc and RRd mean fundamental, first overtone and double mode pulsator, respectively; (3,4) coordinates; (5,6) {\it Gaia} parallax and relative error; (7) ruwe parameter (see text); (8) Period; (9,10,11) $G,G_{BP},G_{RP}$ intensity-averaged magnitudes; (12) number of averaged measures: for APOGEE and GALAH survey it was imposed equal to one; (13,14) adopted iron abundance and relative error; (15) source of the iron abundance: APO=APOGEE; GAL=GALAH; M18=\citet{Magurno2018}.}     
%\label{tab:observationalData}
%\centering
\setlength{\tabcolsep}{2.5pt}
\begin{tabular}{llrrccccccccrcc}
\hline\hline
~~~~~~~~~~~~~~~~~~ID    &    Mode  &     RA    &    Dec   &  $\varpi $ & $\sigma_\varpi$  & ruwe   &  P    &       G  &       GBP   &   GRP &  n & [Fe/H]& $\sigma$[Fe/H] & Source\\
    &          &    (J2000)  & (J2000)   &   (mas) &  (mas)  &        & days  &     (mag)&      (mag)  &  (mag) &  & (dex) &   (dex)     &         \\
~~~~~~~~~~~~~~~~~~(1)    &~~(2)    & (3)    &      (4)    & 	   (5)    &    (6)    &   (7)    &   (8)    & 	  (9)    &    (10)    &  (11)    &   (12)    & (13)    & (14)   & (15) \\	      
\hline
                 V715Cyg      &    RRab  &  295.53335  &   38.91177  &  0.0732  &  0.0460  &  1.055  &   0.47071   &   16.401   &   16.803   &   15.991  &  1  &   $-$1.13  &    0.13   &  M18  \\
                    DXDel      &    RRab  &  311.86821  &   12.46411  &  1.6849  &  0.0323  &  1.049  &   0.47261   &    9.808   &   10.031   &    9.404  &  4  &   $-$0.40  &    0.16   &  M18  \\
                    XZCyg      &    RRab  &  293.12277  &   56.38809  &  1.5713  &  0.0273  &  1.046  &   0.47360   &    9.675   &   99.999   &   99.999  &  2  &   $-$1.55  &    0.10   &  M18  \\
                  V355Lyr      &    RRab  &  283.35799  &   43.15458  &  0.1911  &  0.0225  &  1.280  &   0.47370   &   14.289   &   99.999   &   99.999  &  1  &   $-$1.14  &    0.13   &  M18  \\
                    UUVir      &    RRab  &  182.14595  &   $-$0.45676  &  1.2086  &  0.0810  &  1.125  &   0.47558   &   10.516   &   99.999   &   99.999  &  3  &   $-$0.81  &    0.10   &  M18  \\
                    XZDra      &    RRab  &  287.42763  &   64.85894  &  1.2956  &  0.0239  &  1.064  &   0.47648   &   10.163   &   10.380   &    9.824  &  2  &   $-$0.82  &    0.10   &  M18  \\
  CRTS\_J213829.6-490054        &    RRd   &  324.62358  &  $-$49.01486  &  0.1764  &  0.0241  &  1.221  &   0.47746   &   13.649   &   13.849   &   13.342  &  1  &  $-$0.79   &   0.08    &  GAL  \\
  CRTS\_J134815.9+395403        &    RRab  &  207.06647  &   39.90065  &  0.5696  &  0.0228  &  1.146  &   0.47852   &   11.839   &   12.030   &   11.512  &  1  &  $-$1.90   &   0.22    &  APO  \\
  ASAS\_J202812-4236.2          &    RRab  &  307.05117  &  $-$42.60213  &  0.1689  &  0.0278  &  1.143  &   0.47938   &   13.747   &   13.994   &   13.464  &  1  &  $-$0.83   &   0.08    &  GAL  \\
                     VInd      &    RRab  &  317.87415  &  $-$45.07492  &  1.4972  &  0.0410  &  1.159  &   0.47959   &    9.864   &   10.012   &    9.562  &  2  &   $-$1.46  &    0.16   &  M18  \\
                  V838Cyg      &    RRab  &  288.51629  &   48.19964  &  0.0642  &  0.0196  &  1.320  &   0.48029   &   14.051   &   99.999   &   99.999  &  1  &   $-$1.01  &    0.13   &  M18  \\
  CRTS\_J074506.2+430641        &    RRab  &  116.27630  &   43.11156  &  0.6227  &  0.0378  &  1.093  &   0.48185   &   11.863   &   12.078   &   11.469  &  1  &  $-$0.62   &   0.10    &  APO  \\
                    BRAqr      &    RRab  &  354.63709  &   $-$9.31878  &  0.6489  &  0.0491  &  0.987  &   0.48188   &   11.421   &   99.999   &   99.999  &  1  &   $-$0.69  &    0.13   &  M18  \\
                 V2178Cyg      &    RRab  &  295.02901  &   38.97234  &  0.1398  &  0.0293  &  1.031  &   0.48702   &   15.372   &   15.846   &   14.976  &  1  &   $-$1.66  &    0.13   &  M18  \\
  OGLE-SMC-RRLYR-5992          &    RRab  &   26.84105  &  $-$73.35031  &  0.2347  &  0.0175  &  1.327  &   0.48761   &   13.843   &   14.091   &   13.463  &  1  &  $-$0.41   &   0.09    &  GAL  \\
  ASAS\_J043355-0025.6          &    RRab  &   68.47898  &   $-$0.42552  &  0.4097  &  0.0432  &  1.140  &   0.48768   &   12.627   &   12.843   &   12.189  &  1  &  $-$0.83   &   0.12    &  APO  \\
               KIC6100702      &    RRab  &  282.65722  &   41.42380  &  0.2924  &  0.0115  &  1.118  &   0.48814   &   13.496   &   13.719   &   13.067  &  1  &   $-$0.16  &    0.13   &  M18  \\
                    DHHya      &    RRab  &  135.06169  &   $-$9.77900  &  0.4687  &  0.0402  &  1.166  &   0.48900   &   12.112   &   99.999   &   99.999  &  1  &   $-$1.53  &    0.13   &  M18  \\
                    SZGem      &    RRab  &  118.43102  &   19.27319  &  0.5874  &  0.0418  &  1.296  &   0.50117   &   11.715   &   11.955   &   11.395  &  1  &   $-$1.65  &    0.13   &  M18  \\
                  V450Lyr      &    RRab  &  287.40263  &   43.36387  &  0.0749  &  0.0550  &  1.084  &   0.50460   &   16.608   &   16.790   &   16.278  &  1  &   $-$1.51  &    0.13   &  M18  \\
  ASASSN\_J130646.56-501617.8   &    RRab  &  196.69395  &  $-$50.27161  &  0.2563  &  0.0271  &  1.105  &   0.50815   &   13.746   &   14.106   &   13.256  &  1  &  $-$0.52   &   0.09    &  GAL  \\
  ZTF\_J204705.43-091908.4      &    RRab  &  311.77265  &   $-$9.31905  &  0.4324  &  0.0395  &  1.144  &   0.50875   &   12.541   &   12.811   &   12.150  &  1  &  $-$0.40   &   0.06    &  GAL  \\
                    VWScl      &    RRab  &   19.56251  &  $-$39.21262  &  0.8497  &  0.0727  &  1.517  &   0.51092   &   11.018   &   11.154   &   10.705  &  1  &   $-$1.22  &    0.13   &  M18  \\
                    ANSer      &    RRab  &  238.37939  &   12.96110  &  0.9510  &  0.0446  &  1.134  &   0.52206   &   10.847   &   11.051   &   10.494  &  1  &    0.05  &    0.13   &  M18  \\
                  V782Cyg      &    RRab  &  297.82079  &   40.44586  &  0.2116  &  0.0241  &  1.069  &   0.52364   &   15.265   &   15.727   &   14.627  &  1  &   $-$0.42  &    0.13   &  M18  \\
                  V366Lyr      &    RRab  &  287.41933  &   46.28834  &  0.0349  &  0.0377  &  1.007  &   0.52704   &   16.383   &   16.612   &   15.908  &  1  &   $-$1.16  &    0.13   &  M18  \\
                    FNLyr      &    RRab  &  287.59277  &   42.45883  &  0.2978  &  0.0285  &  1.197  &   0.52740   &   12.652   &   12.971   &   12.300  &  1  &   $-$1.98  &    0.13   &  M18  \\
                    TWBoo      &    RRab  &  221.27477  &   41.02870  &  0.7238  &  0.0228  &  1.075  &   0.53226   &   11.178   &   11.361   &   10.847  &  1  &   $-$1.47  &    0.13   &  M18  \\
                    DOVir      &    RRab  &  219.69150  &   $-$5.32539  &  0.2366  &  0.0392  &  1.096  &   0.53279   &   13.973   &   14.186   &   13.626  &  1  &   $-$1.57  &    0.13   &  M18  \\
               KIC9658012      &    RRab  &  295.33334  &   46.39128  &  0.0599  &  0.0285  &  1.102  &   0.53318   &   15.708   &   16.002   &   15.307  &  1  &   $-$1.28  &    0.13   &  M18  \\
                  V784Cyg      &    RRab  &  299.09543  &   41.33982  &  0.1599  &  0.0299  &  0.972  &   0.53408   &   15.593   &   16.137   &   14.949  &  1  &   $-$0.05  &    0.13   &  M18  \\
  CRTS\_J174421.1-634728        &    RRab  &  266.08802  &  $-$63.79115  &  0.1369  &  0.0251  &  1.054  &   0.53773   &   13.714   &   13.978   &   13.351  &  1  &  $-$0.55   &   0.07    &  GAL  \\
                    UVOct      &    RRab  &  248.10400  &  $-$83.90343  &  1.8925  &  0.0278  &  1.080  &   0.54259   &    9.225   &    9.782   &    9.022  &  2  &   $-$1.75  &    0.10   &  M18  \\
  ASASSN\_J131525.38-752744.2   &    RRab  &  198.85566  &  $-$75.46227  &  0.2584  &  0.0182  &  0.943  &   0.54488   &   14.076   &   14.419   &   13.554  &  1  &  $-$0.40   &   0.08    &  GAL  \\
                  V808Cyg      &    RRab  &  296.41260  &   39.51485  &  0.0773  &  0.0305  &  1.007  &   0.54780   &   15.302   &   15.682   &   14.849  &  1  &   $-$1.19  &    0.13   &  M18  \\
                 V2470Cyg      &    RRab  &  289.99148  &   46.88922  &  0.2571  &  0.0155  &  1.103  &   0.54859   &   13.389   &   99.999   &   99.999  &  1  &   $-$0.59  &    0.13   &  M18  \\
  ASAS\_J220237+0342.3          &    RRab  &  330.65461  &    3.70444  &  0.2468  &  0.0420  &  1.189  &   0.54958   &   13.026   &   13.357   &   12.666  &  1  &  $-$0.86   &   0.08    &  GAL  \\
                    BKTuc      &    RRab  &  352.38917  &  $-$72.54446  &  0.3144  &  0.0247  &  1.357  &   0.55006   &   12.727   &   12.883   &   12.342  &  1  &   $-$1.65  &    0.13   &  M18  \\
  CRTS\_J201956.7-450726        &    RRab  &  304.98684  &  $-$45.12391  &  0.1612  &  0.0267  &  1.172  &   0.55041   &   13.986   &   14.206   &   13.649  &  1  &  $-$1.06   &   0.09    &  GAL  \\
                     WCVn      &    RRab  &  211.61649  &   37.82811  &  1.0185  &  0.0396  &  1.111  &   0.55174   &   10.436   &   10.687   &   10.101  &  1  &   $-$1.18  &    0.13   &  M18  \\
  ASAS\_J213804-4441.2          &    RRab  &  324.51483  &  $-$44.68671  &  0.5043  &  0.0362  &  1.019  &   0.55245   &   12.367   &   12.605   &   11.992  &  1  &  $-$0.21   &   0.07    &  GAL  \\
                    RRCet      &    RRab  &   23.03410  &    1.34154  &  1.5192  &  0.0763  &  1.014  &   0.55304   &    9.616   &    9.935   &    9.314  &  6  &   $-$1.41  &    0.14   &  M18  \\
                    ASVir      &    RRab  &  193.19118  &  $-$10.26028  &  0.5679  &  0.0373  &  1.147  &   0.55345   &   11.843   &   12.078   &   11.500  &  2  &   $-$1.68  &    0.11   &  M18  \\
                  V353Lyr      &    RRab  &  283.00766  &   45.30876  &  0.1034  &  0.0591  &  1.011  &   0.55684   &   16.965   &   99.999   &   99.999  &  1  &   $-$1.50  &    0.13   &  M18  \\
               KIC9717032      &    RRab  &  294.57981  &   46.46302  &  0.1753  &  0.0582  &  0.996  &   0.55691   &   16.828   &   17.218   &   16.414  &  1  &   $-$1.27  &    0.13   &  M18  \\
                  V360Lyr      &    RRab  &  285.49432  &   46.44601  &  0.0113  &  0.0355  &  0.971  &   0.55757   &   16.008   &   99.999   &   99.999  &  1  &   $-$1.50  &    0.13   &  M18  \\
  CRTS\_J111536.5-423619        &    RRab  &  168.90316  &  $-$42.60480  &  0.2427  &  0.0215  &  1.239  &   0.55786   &   13.167   &   13.542   &   12.770  &  1  &  $-$1.15   &   0.09    &  GAL  \\
                  V354Lyr      &    RRab  &  283.20981  &   41.56371  &  0.0584  &  0.0470  &  1.054  &   0.56170   &   16.137   &   16.409   &   15.737  &  1  &   $-$1.44  &    0.13   &  M18  \\
                    DRAnd      &    RRab  &   16.29478  &   34.21838  &  0.5179  &  0.0377  &  1.160  &   0.56312   &   12.352   &   12.613   &   12.002  &  1  &   $-$1.37  &    0.13   &  M18  \\
  ASAS\_J213609-7718.2          &    RRab  &  324.03878  &  $-$77.30388  &  0.5549  &  0.0231  &  1.184  &   0.56345   &   11.953   &   12.222   &   11.597  &  1  &  $-$1.22   &   0.07    &  GAL  \\
                 V1107Cyg      &    RRab  &  289.93865  &   47.10123  &  0.1355  &  0.0365  &  1.059  &   0.56580   &   15.976   &   16.216   &   15.646  &  1  &   $-$1.29  &    0.13   &  M18  \\
  OGLE-SMC-RRLYR-6023          &    RRab  &   27.44191  &  $-$66.24675  &  0.2339  &  0.0144  &  1.127  &   0.56744   &   13.624   &   13.844   &   13.239  &  1  &  $-$0.42   &   0.10    &  GAL  \\
  OGLE-SMC-RRLYR-2614          &    RRab  &  357.63569  &  $-$78.68247  &  0.3811  &  0.0160  &  1.093  &   0.56772   &   12.974   &   13.235   &   12.548  &  1  &  $-$0.86   &   0.08    &  GAL  \\
                    DTHya      &    RRab  &  178.50077  &  $-$31.26111  &  0.2918  &  0.0250  &  1.055  &   0.56798   &   12.959   &   13.166   &   12.569  &  2  &   $-$1.43  &    0.10   &  M18  \\
                    SWDra      &    RRab  &  184.44395  &   69.51059  &  1.0141  &  0.0298  &  1.125  &   0.56967   &   10.386   &   99.999   &   99.999  &  2  &   $-$1.18  &    0.10   &  M18  \\
                    TYGru      &    RRab  &  334.16429  &  $-$39.93833  &  0.1813  &  0.0388  &  1.037  &   0.57002   &   14.021   &   14.327   &   13.734  &  1  &   $-$1.99  &    0.13   &  M18  \\
                    RVOct      &    RRab  &  206.63074  &  $-$84.40171  &  1.0115  &  0.0293  &  1.147  &   0.57116   &   10.841   &   11.104   &   10.384  &  2  &   $-$1.50  &    0.10   &  M18  \\
                  V894Cyg      &    RRab  &  293.25375  &   46.23974  &  0.2391  &  0.0172  &  1.304  &   0.57138   &   12.990   &   13.291   &   12.663  &  1  &   $-$1.66  &    0.13   &  M18  \\
                    CDVel      &    RRab  &  146.15914  &  $-$45.87685  &  0.5654  &  0.0269  &  1.138  &   0.57352   &   11.898   &   12.166   &   11.488  &  2  &   $-$1.78  &    0.10   &  M18  \\
                    RXCet      &    RRab  &    8.40938  &  $-$15.48770  &  0.6483  &  0.0782  &  0.997  &   0.57362   &   11.272   &   11.365   &   10.880  &  1  &   $-$1.38  &    0.13   &  M18  \\
\hline
\end{tabular}
\end{table*}

\begin{table*}
\contcaption{}
%\caption{Sample of RR Lyrae stars used for the comparison with the models with metallicity from high resolution spectroscopy additional with respect to \citet{Magurno2018}. The meaning of the columns is: (1) {\it Gaia} source identifiers; (2) mode of pulsation: RRab, RRc and RRd mean fundamental, first overtone and double mode pulsator, respectively; (3,4) coordinates; (5,6) {\it Gaia} parallax and relative error; (7) ruwe parameter (see text); (8) Period; (9,10,11) $G,G_{BP},G_{RP}$ intensity-averaged magnitudes; (12) number of averaged measures: for APOGEE and GALAH survey it was imposed equal to one; (13,14) adopted iron abundance and relative error; (15) source of the iron abundance: APO=APOGEE; GAL=GALAH; M18=\citet{Magurno2018}.}     
%\label{tab:observationalData}
%\centering
\setlength{\tabcolsep}{2.5pt}
\begin{tabular}{llrrccccccccrcc}
\hline\hline
~~~~~~~~~~~~~~~~~~ID    &    Mode  &     RA    &    Dec   &  $\varpi $ & $\sigma_\varpi$  & ruwe   &  P    &       G  &       GBP   &   GRP &  n & [Fe/H]& $\sigma$[Fe/H] & Source\\
    &          &    (J2000)  & (J2000)   &   (mas) &  (mas)  &        & days  &     (mag)&      (mag)  &  (mag) &  & (dex) &   (dex)     &         \\
~~~~~~~~~~~~~~~~~~(1)    &~~(2)    & (3)    &      (4)    & 	   (5)    &    (6)    &   (7)    &   (8)    & 	  (9)    &    (10)    &  (11)    &   (12)    & (13)    & (14)   & (15) \\	      
\hline
 ASAS\_J201425-5255.7          &    RRab  &  303.60489  &  $-$52.92803  &  0.2906  &  0.0330  &  1.103  &   0.57543   &   13.157   &   13.422   &   12.750  &  1  &  $-$0.60   &   0.09    &  GAL  \\
  CRTS\_J154651.6-380040        &    RRab  &  236.71494  &  $-$38.01139  &  0.4972  &  0.0330  &  1.039  &   0.57606   &   12.801   &   13.068   &   12.215  &  1  &  $-$0.30   &   0.06    &  GAL  \\
                    IOLyr      &    RRab  &  275.65821  &   32.95920  &  0.6487  &  0.0261  &  1.124  &   0.57715   &   11.699   &   11.905   &   11.291  &  1  &   $-$1.35  &    0.13   &  M18  \\
                    BSAps      &    RRab  &  245.21454  &  $-$71.67109  &  0.5357  &  0.0254  &  1.079  &   0.58257   &   12.063   &   12.298   &   11.661  &  2  &   $-$1.49  &    0.10   &  M18  \\
                     ZMic      &    RRab  &  319.09467  &  $-$30.28421  &  0.8148  &  0.0657  &  1.023  &   0.58693   &   11.472   &   99.999   &   99.999  &  2  &   $-$1.51  &    0.10   &  M18  \\
                    UVVir      &    RRab  &  185.31961  &    0.36740  &  0.5618  &  0.0514  &  1.443  &   0.58706   &   11.825   &   12.029   &   11.488  &  1  &   $-$1.10  &    0.13   &  M18  \\
                    RXEri      &    RRab  &   72.43448  &  $-$15.74122  &  1.6232  &  0.0307  &  1.032  &   0.58725   &    9.555   &    9.842   &    9.171  &  1  &   $-$1.18  &    0.13   &  M18  \\
                    XZAps      &    RRab  &  223.02215  &  $-$79.67955  &  0.4267  &  0.0270  &  1.232  &   0.58726   &   12.182   &   12.511   &   11.828  &  2  &   $-$1.78  &    0.13   &  M18  \\
                    NQLyr      &    RRab  &  286.95158  &   42.29853  &  0.2972  &  0.0159  &  1.290  &   0.58778   &   13.296   &   13.517   &   12.911  &  1  &   $-$1.89  &    0.13   &  M18  \\
  ASAS\_J125948-5015.8          &    RRab  &  194.94908  &  $-$50.26295  &  0.3121  &  0.0268  &  1.241  &   0.58828   &   13.325   &   13.691   &   12.791  &  1  &  $-$0.97   &   0.08    &  GAL  \\
  OGLE-LMC-RRLYR-24915         &    RRab  &   42.90222  &  $-$72.17113  &  0.2100  &  0.0160  &  1.242  &   0.58831   &   13.621   &   13.866   &   13.243  &  1  &  $-$0.88   &   0.09    &  GAL  \\
  ASAS\_J173423-6908.5          &    RRab  &  263.59283  &  $-$69.14114  &  0.1850  &  0.0176  &  1.156  &   0.58872   &   13.695   &   13.946   &   13.265  &  1  &  $-$0.75   &   0.09    &  GAL  \\
  ASAS\_J081624-1513.4          &    RRab  &  124.09838  &  $-$15.22275  &  0.3576  &  0.0237  &  1.337  &   0.58972   &   13.194   &   13.456   &   12.883  &  1  &  $-$0.19   &   0.08    &  GAL  \\
                  V350Lyr      &    RRab  &  282.28484  &   46.19860  &  0.0579  &  0.0301  &  1.085  &   0.59424   &   15.678   &   15.996   &   15.381  &  1  &   $-$1.83  &    0.13   &  M18  \\
  ASAS\_J194745-4539.7          &    RRab  &  296.93725  &  $-$45.66052  &  0.1020  &  0.0237  &  1.015  &   0.59633   &   13.353   &   13.514   &   13.028  &  1  &  $-$1.10   &   0.09    &  GAL  \\
                    TTLyn      &    RRab  &  135.78194  &   44.58541  &  1.2192  &  0.0363  &  1.049  &   0.59744   &    9.742   &    9.847   &    9.370  &  3  &   $-$1.50  &    0.10   &  M18  \\
  GDR2\_6082309213159388032     &    RRab  &  202.55332  &  $-$49.79140  &  0.1981  &  0.0257  &  1.230  &   0.60002   &   13.882   &   14.194   &   13.470  &  1  &  $-$1.06   &   0.08    &  GAL  \\
                    SXFor      &    RRab  &   52.59331  &  $-$36.05378  &  0.7554  &  0.0247  &  1.028  &   0.60535   &   10.994   &   11.219   &   10.623  &  1  &   $-$1.80  &    0.13   &  M18  \\
                    UUCet      &    RRab  &    1.02150  &  $-$16.99767  &  0.3823  &  0.0440  &  1.025  &   0.60610   &   11.963   &   12.167   &   11.569  &  3  &   $-$1.35  &    0.10   &  M18  \\
  ASAS\_J210137-3539.2          &    RRab  &  315.40242  &  $-$35.65185  &  0.2227  &  0.0275  &  1.112  &   0.60684   &   13.227   &   13.514   &   12.798  &  1  &  $-$0.63   &   0.08    &  GAL  \\
              KIC11125706      &    RRab  &  285.24486  &   48.74502  &  0.6178  &  0.0198  &  0.917  &   0.61322   &   11.701   &   11.911   &   11.309  &  1  &   $-$1.09  &    0.13   &  M18  \\
                    AEDra      &    RRab  &  276.77804  &   55.49246  &  0.3356  &  0.0214  &  1.240  &   0.61450   &   12.471   &   12.663   &   12.118  &  1  &   $-$1.46  &    0.13   &  M18  \\
                    ATAnd      &    RRab  &  355.62841  &   43.01413  &  2.1779  &  0.2715  &  9.344  &   0.61691   &   10.514   &   10.835   &   10.037  &  1  &   $-$1.41  &    0.13   &  M18  \\
                  V783Cyg      &    RRab  &  298.21977  &   40.79320  &  0.1873  &  0.0219  &  0.938  &   0.62070   &   14.687   &   15.072   &   14.131  &  1  &   $-$1.16  &    0.13   &  M18  \\
                    STBoo      &    RRab  &  232.66340  &   35.78448  &  0.7386  &  0.0525  &  1.077  &   0.62227   &   10.929   &   11.229   &   10.658  &  3  &   $-$1.63  &    0.10   &  M18  \\
  GDR2\_5820212372985755648     &    RRab  &  234.62581  &  $-$69.10700  &  0.2790  &  0.0255  &  1.000  &   0.62246   &   13.037   &   13.322   &   12.583  &  1  &  $-$0.63   &   0.07    &  GAL  \\
                    SSLeo      &    RRab  &  173.47695  &   $-$0.03345  &  0.6859  &  0.0524  &  1.169  &   0.62635   &   11.017   &   99.999   &   99.999  &  2  &   $-$1.91  &    0.43   &  M18  \\
  ASASSN\_J205201.97-791330.2   &    RRab  &  313.00823  &  $-$79.22510  &  0.2417  &  0.0164  &  1.050  &   0.62738   &   13.705   &   14.047   &   13.188  &  1  &  $-$0.99   &   0.09    &  GAL  \\
  GDR2\_1111846056593351168     &    RRab  &  111.86633  &   72.70335  &  1.6251  &  0.0270  &  1.008  &   0.62840   &    9.364   &    9.626   &    8.993  &  1  &  $-$1.93   &   0.16    &  APO  \\
  CRTS\_J214753.4-471332        &    RRab  &  326.97277  &  $-$47.22557  &  0.2407  &  0.0227  &  1.159  &   0.63250   &   13.717   &   13.899   &   13.327  &  1  &  $-$0.87   &   0.09    &  GAL  \\
  CRTS\_J152452.7-284321        &    RRab  &  231.21980  &  $-$28.72222  &  0.2567  &  0.0271  &  0.948  &   0.63290   &   13.606   &   13.980   &   13.070  &  1  &  $-$0.36   &   0.09    &  GAL  \\
  ASAS\_J190237-5636.7          &    RRab  &  285.65200  &  $-$56.61068  &  0.3122  &  0.0500  &  1.036  &   0.63822   &   13.092   &   13.377   &   12.634  &  1  &  $-$1.07   &   0.09    &  GAL  \\
  CRTS\_J222314.6+064802        &    RRab  &  335.81121  &    6.80071  &  0.2645  &  0.0360  &  1.163  &   0.63935   &   13.910   &   14.161   &   13.491  &  1  &  $-$0.73   &   0.08    &  GAL  \\
                     WTuc      &    RRab  &   14.54050  &  $-$63.39574  &  0.5657  &  0.0256  &  0.958  &   0.64225   &   11.351   &   11.638   &   11.063  &  1  &   $-$1.76  &    0.13   &  M18  \\
  ASAS\_J162450+0804.2          &    RRab  &  246.20687  &    8.07054  &  0.2901  &  0.0261  &  1.188  &   0.64473   &   13.023   &   13.266   &   12.641  &  1  &  $-$1.26   &   0.08    &  GAL  \\
  CRTS\_J203652.4-391206        &    RRab  &  309.21794  &  $-$39.20164  &  0.1887  &  0.0294  &  1.126  &   0.65446   &   14.096   &   14.333   &   13.679  &  1  &  $-$0.77   &   0.09    &  GAL  \\
  NSPav                        &    RRab  &  315.71789  &  $-$74.32872  &  0.2644  &  0.0170  &  1.078  &   0.65751   &   13.450   &   13.730   &   13.013  &  1  &  $-$0.99   &   0.09    &  GAL  \\
                    SUDra      &    RRab  &  174.48535  &   67.32940  &  1.4016  &  0.0308  &  1.090  &   0.66041   &    9.663   &    9.884   &    9.365  &  2  &   $-$1.77  &    0.10   &  M18  \\
           BPSCS22881-039      &    RRab  &  332.39762  &  $-$40.43092  &  0.0483  &  0.0376  &  1.001  &   0.66870   &   14.786   &   15.028   &   14.414  &  1  &   $-$2.75  &    0.13   &  M18  \\
                    NRLyr      &    RRab  &  287.11337  &   38.81279  &  0.3156  &  0.0206  &  1.064  &   0.68201   &   12.605   &   12.898   &   12.190  &  1  &   $-$2.54  &    0.13   &  M18  \\
               KIC7030715      &    RRab  &  290.85213  &   42.52837  &  0.2469  &  0.0180  &  1.233  &   0.68361   &   13.129   &   13.394   &   12.698  &  1  &   $-$1.33  &    0.13   &  M18  \\
                    AWDra      &    RRab  &  285.19994  &   50.09196  &  0.2989  &  0.0193  &  1.023  &   0.68718   &   12.712   &   12.895   &   12.353  &  1  &   $-$1.33  &    0.13   &  M18  \\
                    UZCVn      &    RRab  &  187.61539  &   40.50879  &  0.5146  &  0.0329  &  1.167  &   0.69779   &   11.994   &   12.209   &   11.631  &  1  &   $-$2.21  &    0.13   &  M18  \\
  GDR2\_4330459501285066496     &    RRab  &  248.19958  &  $-$13.09923  &  0.0491  &  0.0403  &  1.068  &   0.70255   &   15.221   &   15.717   &   14.542  &  1  &  $-$1.51   &   0.13    &  APO  \\
  CRTS\_J121320.4-234334        &    RRab  &  183.33495  &  $-$23.72620  &  0.1665  &  0.0312  &  1.111  &   0.71003   &   13.799   &   14.031   &   13.385  &  1  &  $-$1.30   &   0.09    &  GAL  \\
  ASASSN\_J185553.56-664412.0   &    RRab  &  283.97311  &  $-$66.73678  &  0.2293  &  0.0190  &  1.280  &   0.71960   &   13.304   &   13.536   &   12.938  &  1  &  $-$0.85   &   0.09    &  GAL  \\
  CRTS\_J123842.8-291327        &    RRab  &  189.67803  &  $-$29.22421  &  0.2263  &  0.0271  &  1.212  &   0.77823   &   13.597   &   13.818   &   13.197  &  1  &  $-$1.01   &   0.08    &  GAL  \\
  OGLE-SMC-RRLYR-2715          &    RRab  &    0.50096  &  $-$76.55434  &  0.2341  &  0.0183  &  1.209  &   0.81813   &   13.138   &   13.411   &   12.700  &  1  &  $-$1.26   &   0.07    &  GAL  \\
\hline
\end{tabular}
\end{table*}

%%%%%%%%%%%%%%%%%%%%%%%%%%%%%%%%%%%%%%%%%%%%%%%%%%

% Don't change these lines
\bsp	% typesetting comment
\label{lastpage}
\end{document}